\documentclass[twocolumn,apj,numberedappendix,appendixfloats,twocolappendix]{openjournal}

\usepackage{newtxtext,newtxmath}

\usepackage[T1]{fontenc}

\usepackage{graphicx}
\usepackage[colorlinks,linkcolor=blue,citecolor=blue,urlcolor=blue ]{hyperref}
\usepackage{amsmath}
\usepackage{soul}
\usepackage{xspace}
\usepackage{ulem}                   
\usepackage[dvipsnames]{xcolor}
\usepackage{pifont}
\usepackage{orcidlink}
\usepackage{tabularx}

\makeatletter 
  
  \patchcmd{\NAT@citex}
    {\@citea\NAT@hyper@{%
      \NAT@nmfmt{\NAT@nm}%
      \hyper@natlinkbreak{\NAT@aysep\NAT@spacechar}{\@citeb\@extra@b@citeb}%
      \NAT@date}}
    {\@citea\NAT@nmfmt{\NAT@nm}%
    \NAT@aysep\NAT@spacechar\NAT@hyper@{\NAT@date}}{}{}

  \patchcmd{\NAT@citex}
    {\@citea\NAT@hyper@{%
      \NAT@nmfmt{\NAT@nm}%
      \hyper@natlinkbreak{\NAT@spacechar\NAT@@open\if*#1*\else#1\NAT@spacechar\fi}%
        {\@citeb\@extra@b@citeb}%
      \NAT@date}}
    {\@citea\NAT@nmfmt{\NAT@nm}%
    \NAT@spacechar\NAT@@open\if*#1*\else#1\NAT@spacechar\fi\NAT@hyper@{\NAT@date}}
    {}{}
\makeatother

\DeclareRobustCommand{\ion}[2]{%
\relax\ifmmode
\ifx\testbx\f@series
{\mathbf{#1\,\mathsc{#2}}}\else
{\mathrm{#1\,\mathsc{#2}}}\fi
\else\textup{#1\,{\mdseries\textsc{#2}}}%
\fi}

\newcommand{\LCDM}{$\Lambda$CDM\xspace}

\newcommand{\Msun}{{\rm M_\odot}}

\newcommand{\hMpc}{h^{-1}\,{\rm Mpc}}

\newcommand{\ie}{{i.e.~}}
\newcommand{\eg}{{e.g.~}}

\newcommand{\HI}{\ion{H}{I}\xspace}
\newcommand{\HII}{\ion{H}{II}\xspace}

\newcommand{\HeII}{\ion{He}{II}\xspace}
\newcommand{\HeIII}{\ion{He}{III}\xspace}
\newcommand{\CIV}{$[$\ion{C}{IV}$]\lambda1548,1550$\xspace}

\newcommand{\OIII}{$[$\ion{O}{III}$]\lambda4960,5008$\xspace}

\newcommand{\Lya}{Ly$\alpha$\xspace}
\newcommand{\Lyb}{Ly$\beta$\xspace}
\newcommand{\Lyc}{Ly$\gamma$\xspace}
\newcommand{\Lyd}{Ly$\delta$\xspace}
\newcommand{\taueff}{\tau_\mathrm{eff}}

\newcommand{\citenp}[1]{\citeauthor{#1} \citeyear{#1}}

\newcommand{\pMpc}{\mathrm{pMpc}}

\newcommand{\arepo}{{\sc arepo}\xspace}
\newcommand{\areport}{{\sc arepo-rt}\xspace}

\newcommand{\thesan}         {\textsc{thesan}\xspace}
\newcommand{\thesanone}      {\textsc{thesan-1}\xspace}

\newcommand{\thesanlow}      {\textsc{thesan-low-2}\xspace}
\newcommand{\thesanhigh}     {\textsc{thesan-high-2}\xspace}

\newcommand{\Ltot}{L_\mathrm{tot}}
\newcommand{\galacc}{GaL$\alpha$CC\xspace} 

\graphicspath{{./}{figures/}}

\shorttitle{The galaxy--IGM connection in \thesan pt. 2: the galaxy--Lyman-$\alpha$ cross-correlation}
\shortauthors{Garaldi et al.}

\begin{document}
\title{\textbf{The galaxy--IGM connection in thesan: observability and information content of the galaxy--Lyman-$\boldsymbol{\alpha}$ cross-correlation at z$\geq$6}\vspace{-1.5cm}}
\author{E.~Garaldi\orcidlink{0000-0002-6021-7020},$^{1,2,3,4,5,*,\dagger}$}
\author{V.~Bellscheidt\orcidlink{0009-0006-1543-9907}$^{6}$}
\author{A.~Smith\orcidlink{0000-0002-2838-9033}$^{7}$\vspace{0.2cm}}
\author{R.~Kannan\orcidlink{0000-0001-6092-2187}$^{8}$}
\thanks{$^*$E-mail: \href{mailto:egaraldi@sissa.it}{egaraldi@sissa.it}}
\thanks{$^\dagger$CANON Fellow}
\affiliation{$^{1}$Kavli IPMU (WPI), UTIAS, The University of Tokyo, Kashiwa, Chiba 277-8583, Japan}
\affiliation{$^{2}$Institute for Fundamental Physics of the Universe, via Beirut 2, 34151 Trieste, Italy}
\affiliation{$^{3}$SISSA - International School for Advanced Studies, Via Bonomea 265, 34136 Trieste, Italy}
\affiliation{$^{4}$INAF, Osservatorio Astronomico di Trieste, Via G. B. Tiepolo 11, I-34131 Trieste, Italy}
\affiliation{$^{5}$Department of Earth and Space Science, Osaka University, Toyonaka, Osaka 560-0043, Japan}
\affiliation{$^{6}$Technical University of Munich, TUM School of Natural Sciences, Physics Department, James-Franck-Strasse 1, 85748 Garching, Germany}
\affiliation{$^{7}$Department of Physics, The University of Texas at Dallas, Richardson, TX 75080, USA}
\affiliation{$^{8}$Department of Physics and Astronomy, York University, 4700 Keele Street, Toronto, ON M3J 1P3, Canada}

\begin{abstract}  
\noindent The galaxy--Lyman-$\alpha$ cross-correlation (\galacc) is a promising tool to study the interplay of galaxies and inter-galactic medium (IGM) in the first billion years of the Universe. Here we thoroughly characterise the impact of observational limitations on our ability to retrieve the intrinsic \galacc and provide new physical insights on its origin and connection to other IGM properties. This is extremely relevant to identify promising datasets, design future surveys and assess the limitations of current measurements. 
We find that sightline-to-sightline variations demand at least 25 independent sightlines to quantitatively recover the true signal. Once this condition is met, the intrinsic signal can be recovered even for a relatively low signal-to-noise ratio and spectral resolution. The galaxy selection method does not affect the inferred \galacc and lightcone effects are relevant whenever observations span a redshift window broader than $\Delta z \gtrsim 0.4$. We discuss the implication for previous theoretical studies that did not account for them. We elucidate explicitly for the first time the physical origin of the \galacc and demonstrate that this signal is collectively sourced by the ensemble of galaxies residing in overdense regions rather than individual objects. We show that the \galacc measured for opaque sightlines shows a larger peak at smaller scales with respect to transparent lines of sight. We connect this to the evolution of the mean free path of ionizing photons, showing that the \galacc peak position has a very similar evolution but on smaller scales, as it probes only the core of ionized regions. Finally, we discuss which ongoing surveys can be used to measure the \galacc and provide an initial analysis of future developments, including using galaxies as background sources, and the application to helium reionization. Our results outline a bright future for the \galacc as a tool to unveil the galaxy--IGM interplay during the first billion years of the Universe. 
\end{abstract}
\maketitle

\section{Introduction}

Following the emergence of the first stars and galaxies within a few hundred million years after the Big Bang, the ultraviolet photons emitted by such objects began to ionize the intergalactic medium (IGM) gas between them in what is known as the Epoch of cosmic Reionization (EoR). These ionized regions eventually grew to encompass the entire Universe, marking the end of such epoch.

The study of cosmic reionization is underpinned by a growing body of observational constraints, that just in the last few years started to probe beyond the tail end of this process \citep[for a community-updated collection see \eg \url{corecon.readthedocs.io};][]{corecon}. These include global constraints from the cosmic microwave background \citep[CMB, \eg][]{Planck2018cosmo,Pagano+2020,deBelsunce+2021} and localised (in time and space) constraints through --~among others~-- the Lyman-$\alpha$ (\Lya) absorption in quasar spectra \citep[\eg][]{Fan+2006,McGreer+2011,Yang+2020,Lu+2020,Bosman+2021}, the visibility evolution of galaxies \citep[\eg][]{Ota+2008, Pentericci+2014, Mesinger+2015}, as well as the damping wing of both quasars \citep[\eg][]{Mortlock+2011, Greig+2017, Wang+2020, Durovcikova+2024} and Gunn-Peterson troughs \citep{Spina+2024,Zhu+2024}. 

The launch of the \textit{James Webb Space Telescope} (JWST) has enabled a fast and dramatic progress on the characterisation of the sources of reionization. Thanks to its near-infrared capabilities and sensitivity, it has already unveiled the properties of a large number of `normal' galaxies within the first billion years of the Universe \citep[\eg][]{ceersI, EIGERII, Harikane+2023, jades, fresco}. This is not only extremely important to progress further in our understanding of primeval galaxy formation and cosmic reionization, but allows us for the first time to observationally study how these two processes influence each other in the reionizaing Universe. 

In recent years, the cross-correlation between the transmitted flux in the Lyman-$\alpha$ (\Lya) forest and the position of galaxies around the line of sight (hereafter named galaxy--\Lya cross-correlation, or \galacc) has been used to probe the complex galaxy--IGM interplay during the EoR \citep[starting with ][]{Kakiichi+2018}. This quantity shows two prominent features, namely an excess of transmitted \Lya flux at distances $10 \lesssim r / [\hMpc] \lesssim 30$ from galaxies and a strong suppression of such flux at $r \lesssim 10 \, \hMpc$ \citep{Meyer+2019, Meyer+2020}. The former has been interpreted as a (transverse) proximity effect driven by the ionizing radiation field of the galaxies, while the latter is typically ascribed to the overdensity in which galaxies reside boosting the hydrogen recombination rate. The position and amplitude of the flux excess strongly depends on the progress of reionization \citep{Thesan_igm}, rendering it a powerful tool to constrain the timing of reionization. Little is known beyond this, since theoretical studies of the \galacc are very limited, mainly as a consequence of the simultaneous requirements of $\mathcal{O}(100 \, \mathrm{Mpc})$ scales to properly capture the inhomogeneous reionization process \citep{Iliev+2014, Kaur+2020, GnedinMadau_review} while resolving galactic properties \citep[since a failure to do so can erase the \galacc,][]{Garaldi+2019croc, Thesan_igm}. During the revision process of this manuscript, \citet{Conaboy+2025} published a thorough investigation of the \galacc using the Sherwood-relics simulations \citep{sherwood-relics}, which employ a two-step process to include the impact of gas photo-ionisation and photo-heating in the simulation without performing on-the-fly radiation transport. 

Observationally studying the \galacc is also difficult due to the simultaneous requirements of spectroscopy and galaxy detection around the line of sight. This has so far limited studies of this cross-correlation to a small number of sightlines. The status quo is however rapidly changing thanks to the JWST. Recently, the EIGER program published their measurements of the \galacc in \citet{EIGERI}, based on the first observed line of sight (out of six planned). Intriguingly, matching their results to numerical predictions \citep[from][]{Thesan_igm} requires a very late end of the EoR. Interestingly, this is in contrast to what is inferred by matching the observations of \citet{Meyer+2019} to the simulations in \citet{Conaboy+2025}. The ASPIRE \citep{aspire} program is also measuring the \galacc, thanks to its NIRCam/WFSS observations of 25 quasar fields in the redshift range $6.5 < z_\mathrm{QSO} < 6.8$. Recently, results from the first 5 quasar fields were published \citep{ASPIRE_galacc}, demonstrating an overall good agreement with the \thesan simulations. Nevertheless, a detailed comparison revealed some differences pointing towards larger ionised bubbles and stronger UV background and temperature fluctuations. 
Despite such progress, available observations remain sparse and with significant differences in their galaxy identification method (\eg \CIV absorption, \OIII emission, Lyman-break, etc.), quasar spectrum noise level and resolution, and field coverage. However, the impact of these differences has not yet been investigated, nor has been determined a set of minimal requirements for observations of the \galacc. 

In this paper we provide a thorough characterization of the impact on the inferred \galacc of a broad range of observational limitations and choices. This enables us to assess the reliability and guide the interpretation of current and future studies of this quantity, as well as to determine a set of minimal requirements for future observations of the \galacc. We describe the simulation set used and the production of synthetic observations in Sec.~\ref{sec:methods}. In Sec.~\ref{sec:results_obs}  we present our results concerning the impact of observational limitations on the inferred \galacc, while in Sec.~\ref{sec:results_phys} we connect the properties of the (simulated) signal to other physical properties of the Universe. Finally, in Sec.~\ref{sec:future} we discuss potential future developments of the \galacc, including the prospect of employing ongoing surveys to extend observations of this quantity, and we present concluding remarks in Sec.~\ref{sec:conclusions}.

\begin{figure*}
    \includegraphics[width=\textwidth]{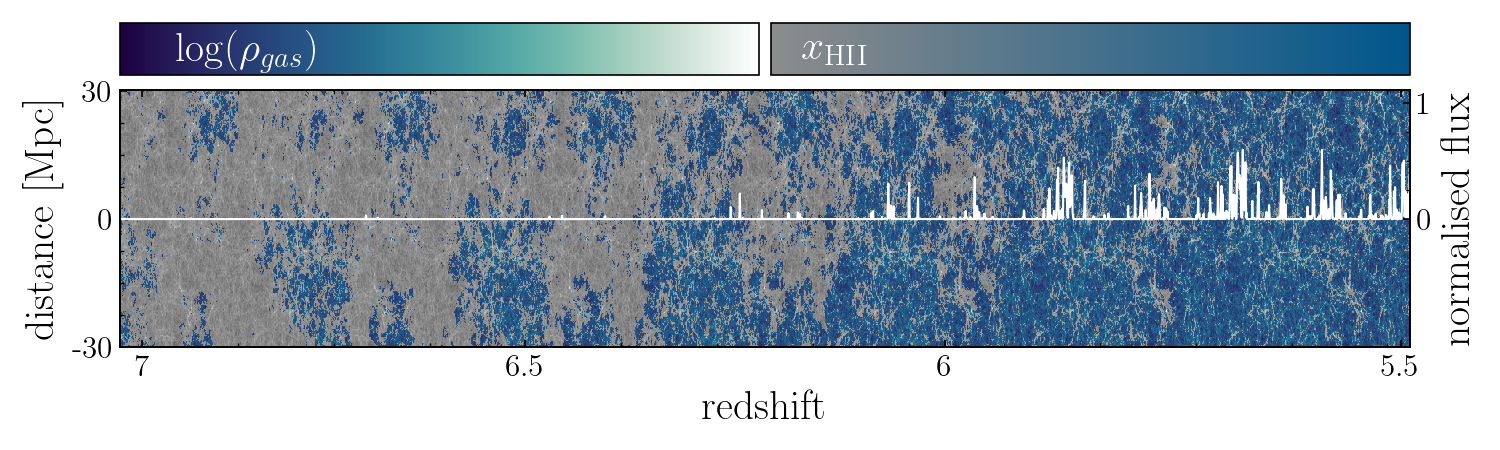}
    \caption{Example of lightcone spectrum (white line) and density field (background map). The spectrum shows the normalised flux in the Lyman-$\alpha$ forest (right vertical axis) at the redshift indicated by the horizontal axis. The background color shows the gas density (in log scale, as indicated by the top left colorbar, where units are arbitrary), while the opacity indicates the \HII fraction ($x_\mathrm{HII}$, in log scale, top right colorbar). The latter is chosen to completely desaturate \HII fractions below $10^{-4}$, that are completely opaque to the Lyman-$\alpha$ photons.
    }
    \label{fig:lc_slice}
\end{figure*}

\section{Methods}
\label{sec:methods}
Faithfully studying the interplay between galaxies and the large-scale reionization of the IGM  is challenging, as it requires simultaneous modeling of sub-galactic and inter-galactic scales in volumes \textit{at least} $V \gtrsim (100 \, \pMpc)^3$ \citep[][but potentially much larger, see \citenp{Iliev+2014}]{GnedinMadau_review}, accounting for the inhomogeneous reionization process and including realistic galaxy formation physics. This is a formidable challenge, that has been conquered only by a handful of simulations to date \citep{CoDa, CoDaII, CoDaIII, CoDaI-AMR, CROC}. In this work we employ one of them, the \thesan suite \citep{Thesan_intro, Thesan_igm, Thesan_Lya, Thesan_data}, which we briefly describe in the following and refer the interested reader to \citet{Thesan_intro} and \citet{Thesan_data} for a more thorough description of the numerical and physical setup, as well as of the numerical products available. 

\subsection{The \thesan simulations}
\label{subsec:thesan}
The \thesan simulation suite is a set of radiation-hydrodynamical simulations, recently made publicly available at \url{www.thesan-project.com} \citep{Thesan_data}. They are designed following two main tenets, \ie (\textit{i}) to simultaneously capture IGM and galactic properties during the EoR while (\textit{ii}) minimizing the number of free parameters at $z\gtrsim5$. In order to do so, the \thesan simulations employ the successful IllustrisTNG galaxy formation model \citep{Weinberger2017, Pillepich2018b}, coupled to the dust model of \citet{McKinnon+16} and the \areport \citep{ArepoRT} radiation transport module of the \arepo code \citep{arepo,Arepo-public}. By maintaining the free parameters of these models fixed to the values calibrated on the low-$z$ Universe, \thesan has a single (additional) free parameter, namely the escape fraction of ionizing photons from the birth cloud of the star.\footnote{This is \textit{not} the escape fraction typically discussed in the context of cosmic reionization, which refers to the ionizing photons escape from the \textit{entire galaxy/halo} and that the simulations can predict \citep{Thesan_fesc}. Rather, this represents the absorption of ionizing photons by unresolved structures around the birth place of stars, typically on scales of $\lesssim 10$~parsecs \citep[see \eg Table 1 of][]{Thesan_data}.} The latter is calibrated by requiring the simulations to approximately match the observed `late' reionization history \citep[\eg][]{Zhu+2020, Bosman+2021, Kulkarni2019, Keating+2020, Nasir&DAloisio2020}.

All \thesan simulations employ a \citet{Planck2015cosmo} cosmology and have a box size of $L_\mathrm{box} = 95.5 \, \mathrm{Mpc}$. The flagship simulation, \thesanone, has a mass resolution sufficient to resolve atomic cooling haloes, the smallest of structures significantly contributing to the ionizing photon budget \citep[for a recent confirmation of the negligible role of mini-haloes see \eg][]{Gnedin2024}. This is the simulation used in this paper. \thesan matches the main galaxy properties observed in the pre-JWST \citep{Thesan_intro} and JWST \citep{Thesan_data, Thesan_sizes} era, as well the observed IGM properties \citep{Thesan_igm, ThesanHR, Thesan_bubbles, Thesan_tracking_bubbles, Thesan_lg}. 

It should be noted that, despite being one of the largest-volume radiation-hydrodynamical simulations of the Universe currently able to capture galaxy properties (only comparable to CROC --~\citealt{CROC}~-- and CoDaIII --~\citealt{CoDaIII}), the volume covered remains somewhat small, and only barely approaching the volumes needed for a converged reionization history \citep[\eg][]{Iliev+2014, Kaur+2020, GnedinMadau_review}. For instance, \citet{Becker+2011} reported the discovery of a very long and very opaque Gunn-Peterson trough in an otherwise mostly ionized Universe. Extreme features like this one can not be captured in our simulations (see \eg \citealt{Keating+2020} for an estimation of the volume needed to simulate similar features). Therefore, it is possible that our results will be marginally affected by a somewhat suppressed variability due to the inability of \thesan to capture the most extreme features. However, we expect these to be very rare, and therefore to bear a small impact on our results.

\subsection{Synthetic spectra}
\label{subsec:spectra}
The main data product used in this paper are synthetic lines of sights (LOS) extracted from the simulation outputs \citep[see Section 3.10 of][]{Thesan_data}. These are extracted using the \textsc{colt} code \citep[last described in][]{Smith+2022}, which uses the native Voronoi tessellation of the simulation to retain the full spatial information available. Using the gas properties extracted in this way, we construct synthetic \Lya forest spectra using the full Voigt-Hjerting line profile (\citenp{Hjerting1938}, through the approximation of \citenp{Harris1948} and \citenp{Tepper-Garcia2006}). We include the effects of gas temperature and peculiar velocities. Our spectra have a spectral resolution of $\Delta \varv = 1 \, \mathrm{km\,s}^{-1}$. For each simulation snapshot (\ie approximately every 11 Myr) having redshift between $5.5\lesssim z \lesssim 7$, we produce 300 lines of sight (in addition to the 150 already available in the public data release). These new LOS originate from a random location in the $xy$ plane, are aligned along the $z$ direction and are $95.5 \, \mathrm{Mpc}$ long. We note here that, thanks to the unstructured mesh of \arepo, the synthetic spectra do not suffer from any grid-alignment artefacts. 

For this paper, we combine the LOS produced at each simulation snapshot to produce lightcone spectra, employing a piecewise-constant approximation. We release these sightlines on the \thesan website (see Appendix~\ref{app:los_on_website} for more information). In Fig.~\ref{fig:lc_slice} we show an example of such lightcone LOS. The background map shows the density distribution within 30 Mpc of the sightline (along the $y$ direction of the simulation, left-hand side vertical axis and top left colorbar). The saturation of this map reflects the amount of neutral hydrogen in the IGM (with grey regions corresponding to $x_\mathrm{HI} = 10^{-4}$ and progressively more saturated colors indicating lower neutral fractions, top right colorbar). The white spectrum super-imposed to this map shows the normalised transmitted flux along the LOS (right-hand side y axis). The evolving redshift of the spectrum is indicated by the horizontal axis. This visually shows the well-known boost in \Lya transmission in highly-ionized gas, which is preferentially found at low redshift. 

\subsection{Computing the \galacc}
\label{subsec:computing_galacc}
Here, we briefly summarize how the \galacc is computed in the paper. Given a set of spectra and a set of galaxy positions, we first compute the distance between each pixel in each spectrum and each galaxy position $d_{i,j,k}$. 
Then, we discretize these distances into bins and compute the average transmitted flux for all pixel-galaxy pairs within each bin. Finally, we divide the result by the average transmitted flux of the entire sample of spectra (\ie using every pixel in every spectrum) and subtract one to adhere to the convention in the literature. 
Notice that, by virtue of the fact that the synthetic spectra are randomly sampling the entire volume and are many in number, this procedure is equivalent to normalizing by the average \Lya transmitted flux in the Universe, as typically done in observations where the number of available spectra is low and the procedure described might yield biased results because of sample variance.

In the paper, when not stated otherwise, the cross-correlation is always computed at $z=6$ using all the 300 sightlines (either at fixed redshift or using the lightcone LOS) and considering galaxies with stellar mass $M_\mathrm{star} \geq 10^8 \, \Msun$ (corresponding to 1834 galaxies in the simulation box at $z=6$). We use 100 bins in the radial direction, linearly-spaced between $0$ and $0.4 L_\mathrm{box} \sqrt{3} / 2 \approx 45\,\hMpc$. Additionally, all quantities are expressed in comoving units, unless specified otherwise.

\section{Impact of observational limitations}
\label{sec:results_obs}

\subsection{How many spectra are necessary to measure the \galacc?}
\label{subsec:how_many}
The first question we address is the following: \textit{How many spectra do we need to obtain statistically-sound results?} The inhomogeneous nature of reionization implies that the gas ionization state and temperature in different regions of the Universe vary significantly until $z\lesssim4$ \citep{Sherwood}. Additionally, the highly non-linear process of galaxy formation, the consequent energy injection (`feedback') into the surrounding circum-galactic medium (CGM) and IGM, as well as the different cosmic environments hosting galaxies of similar mass all can strongly influence the \Lya forest signal. Therefore, individual lines of sight are dominated by cosmic variance (sourced not only by the large-scale structures of the Universe but also by the `structure formation noise' described above). Current observational efforts are still limited to a handful of sightlines, that however might be not sufficient to extract the intrinsic signal. 

\begin{figure*}
    \includegraphics[width=\textwidth]{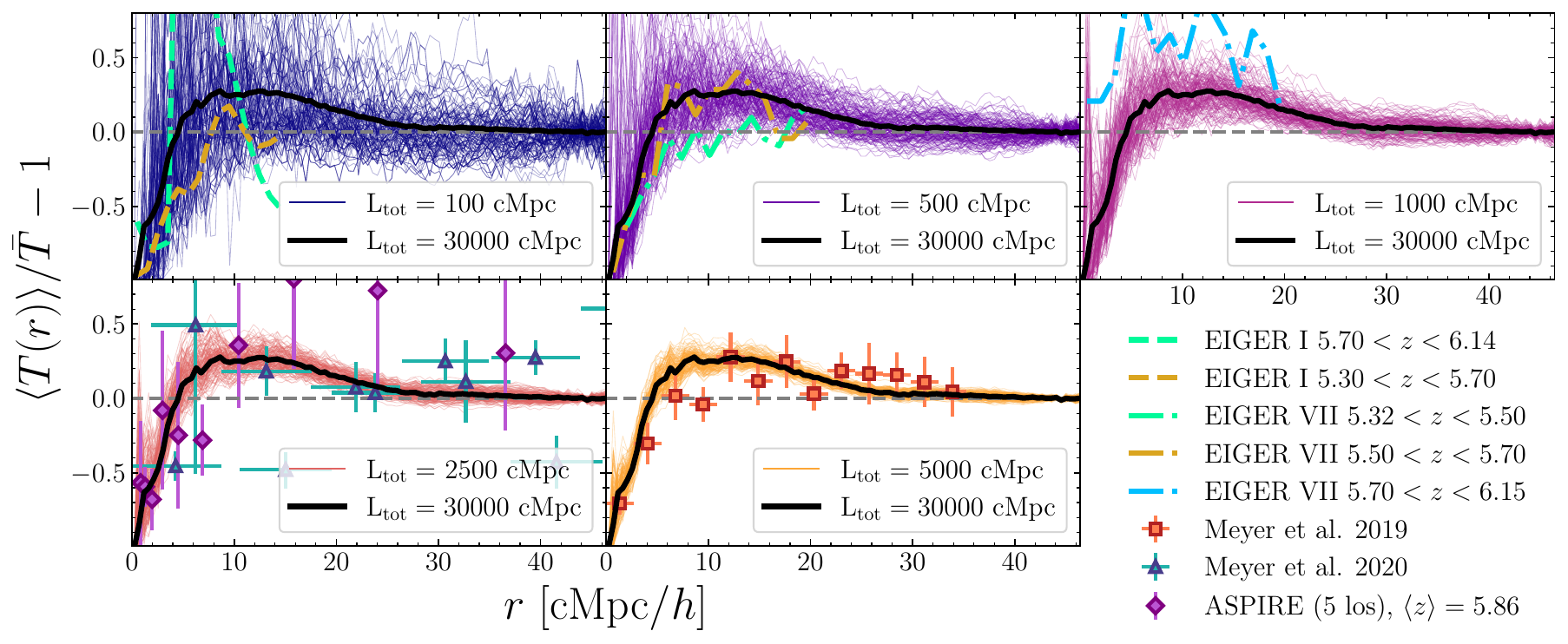}
    \caption{Impact of the total spectral length $\Ltot$ on the prediction of the galaxy--\Lya forest cross-correlation. Each panel shows the signal obtained by averaging \Lya spectra totalling the length reported in the bottom right (colored lines), in comparison with the signal obtained averaging all synthetic spectra produced (corresponding to $\Ltot = 15$ cGpc, black solid line). In the appropriate panel, we show the measurements from \citet[][red squares]{Meyer+2019}, \citet[][blue triangles]{Meyer+2020}, the EIGER JWST program (dashed lines for \citealt{EIGERI}, dot-dashed lines for \citealt{EIGERVII}) and the ASPIRE JWST program \citep[purple diamonds][]{ASPIRE_galacc}. To compute the predicted signal, we have selected all galaxies with stellar mass $M_\mathrm{star} \geq 10^8 \, \Msun$ (corresponding to 1834 galaxies in the simulation box).}
    \label{fig:corr_vary_Llos}
\end{figure*}

\begin{figure*}
    \includegraphics[width=\textwidth]{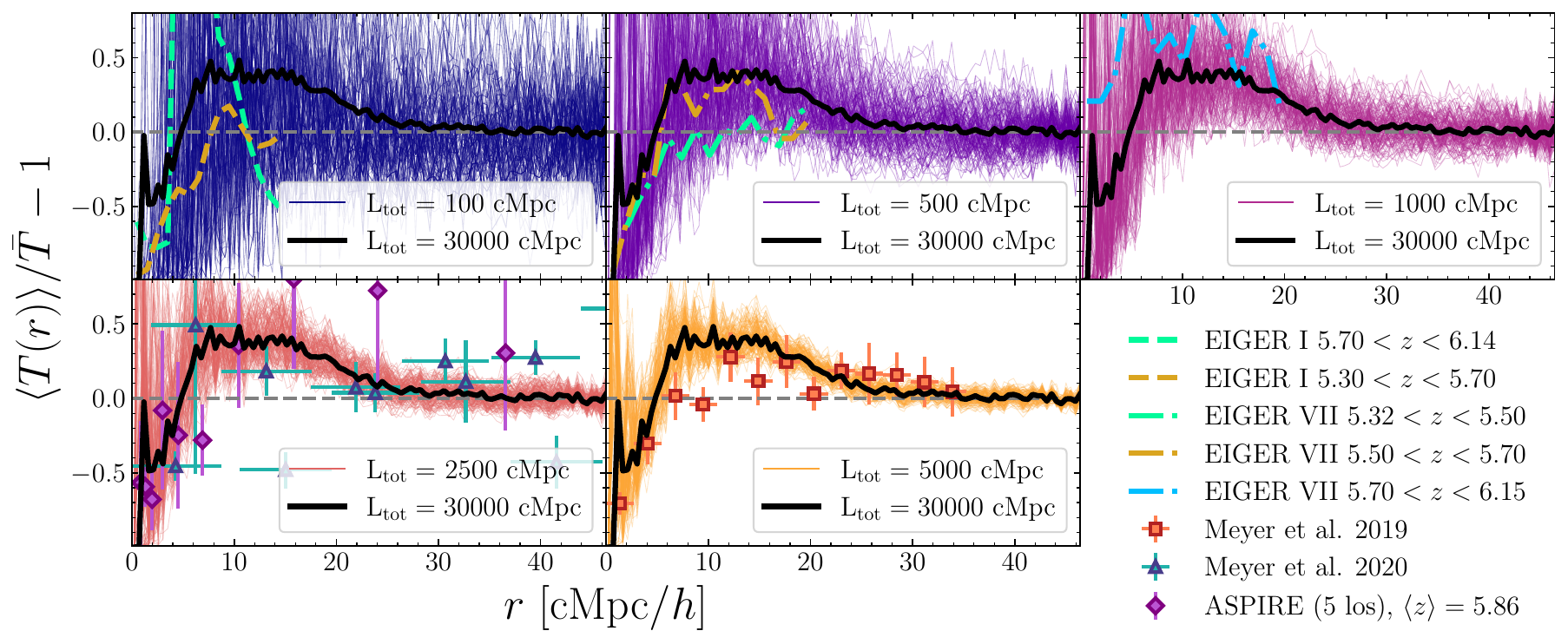}
    \caption{As Fig.~\ref{fig:corr_vary_Llos}, but employing all galaxies with stellar mass $M_\mathrm{star} \geq 10^9 \, \Msun$ (corresponding to 126 galaxies in the simulation box).}
    \label{fig:corr_vary_Llos_Mstar1e9}
\end{figure*}

\begin{figure*}
    \includegraphics[width=\textwidth]{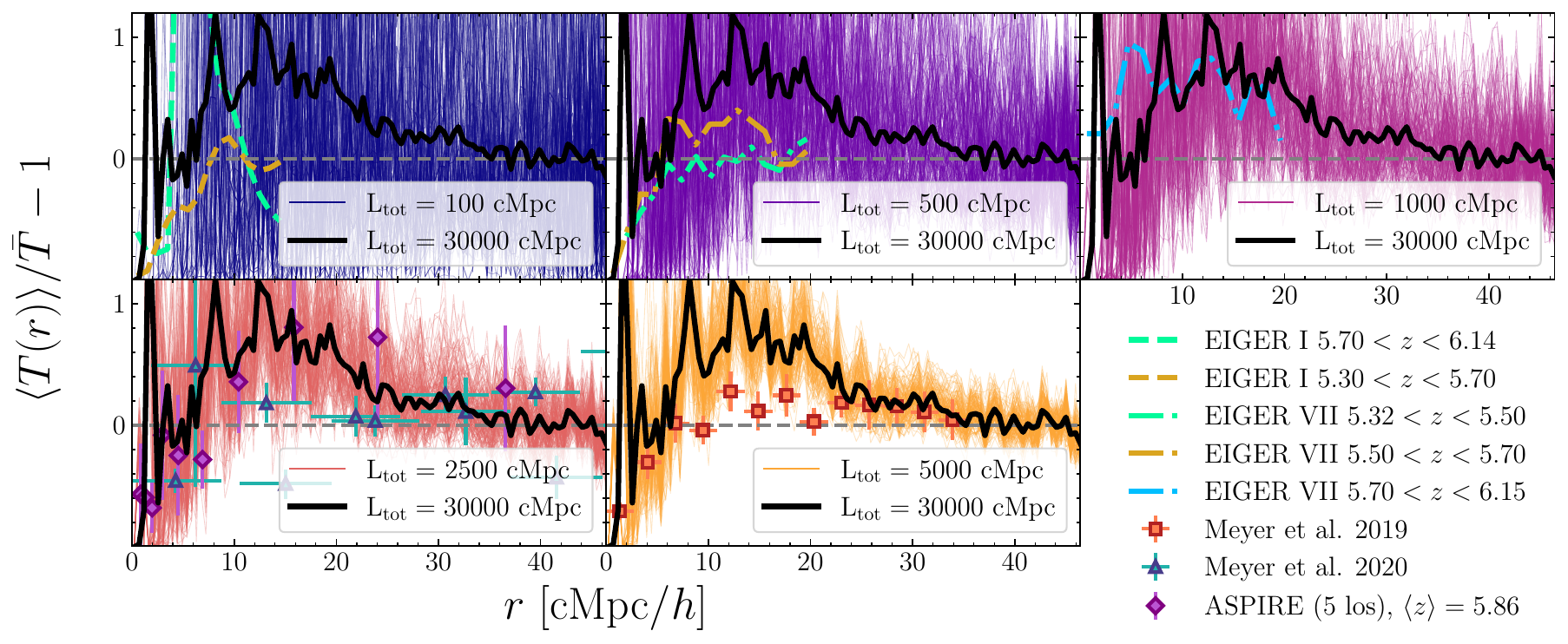}
    \caption{As Fig.~\ref{fig:corr_vary_Llos}, but employing all galaxies with stellar mass $M_\mathrm{star} \geq 10^{10} \, \Msun$ (corresponding to 8 galaxies in the simulation box).}
    \label{fig:corr_vary_Llos_Mstar1e10}
\end{figure*}

In order to test the reliability of current constraints and to guide future observations, we show in Fig.~\ref{fig:corr_vary_Llos} the \galacc signal extracted from our simulations for different values of the total length of the \Lya forest spectra ($\Ltot$). In practice, we employ up to 300 spectra of equal length $L_\mathrm{spec} = 95.5 \, \mathrm{cMpc}$, but decide to report the results using the total spectral length in order to ease the comparison with observations. In each panel, the thin colored lines are the values obtained using different sets of sightlines with total spectral length equal to the value reported in the bottom right part of the panel. For comparison, we show using a thick black line our `golden standard', \ie the value obtained using all 300 spectra available. It appears immediately clear that this signal is cosmic variance-dominated in the top panels (corresponding to $\Ltot \lesssim 1 \, \mathrm{cGpc}$), while the recovered signal becomes close to the intrinsic one (that we assume is traced by our black thick line) for $\Ltot \gtrsim 2.5 \, \mathrm{cGpc}$. 

We compare the prediction from our simulation with available observations, namely \citet[][$\Ltot \approx 12$ cGpc]{Meyer+2019}, \citet[][$\Ltot \approx 3.25$ cGpc]{Meyer+2020}, the EIGER JWST program (dashed lines for \citealt{EIGERI} -- $\Ltot \approx 189$ cMpc --, dot-dashed lines for \citealt{EIGERVII} -- $\Ltot \approx 500 - 1000$ cMpc) and the ASPIRE JWST program \citep[purple diamonds][]{ASPIRE_galacc}. 
Observations from \citet{Meyer+2019,Meyer+2020} and ASPIRE span a large enough path length to overcome cosmic variance and extract the true reionization signal. 
It should be noted, however, that to reach such large $\Ltot$, they employ somewhat shallow observations, reducing the number of galaxies available, and compound together large redshift intervals, potentially encountering lightcone effects due to the rapid evolution of the ionization field towards the end of the EoR (as discussed in Sec.~\ref{subsec:lightcone_effects}). 

In the case of \citet{EIGERI}, reporting the results from a single sightline from the EIGER survey, we predict that their results are completely dominated by cosmic variance at $r \gtrsim 3\,\hMpc$, and therefore cannot be quantitatively nor qualitatively trusted as representative of the average \galacc. 
The aforementioned results put the findings of \citet{EIGERI} in a different light. In fact, while they find a qualitative similarity between their measured $T(r)/\bar{T}-1$ and the one reported by the \thesan simulations at redshift $z \sim 6.7$, such agreement is likely entirely driven by cosmic variance. 
The latest EIGER results \citep{EIGERVII}, instead, include all 6 sightlines and, therefore, correspond to a much larger $\Ltot$. This, combined with their deep observations resulting in a rich galaxy sample, makes them much more reliable in a statistical sense. At smaller distances, the flux suppression due to the overdensities harbouring galaxies dominates over cosmic variance and, therefore, the observed signal can be trusted, at least qualitatively. These results are in much better agreement with numerical predictions and present a clear picture of the transition from a galaxy-dominated radiation field to a UVBG-dominated one.

It should be noted that quantitative results depend on the galaxies selected when computing the cross-correlation (or, similarly, detected by a survey). In particular, in Fig.~\ref{fig:corr_vary_Llos} we have selected all galaxies with stellar mass $M_\mathrm{star} \geq 10^8 \, \Msun$. We provide a visual impression of the impact of such choice in Fig.~\ref{fig:corr_vary_Llos_Mstar1e9} and \ref{fig:corr_vary_Llos_Mstar1e10}, where we have selected all galaxies with stellar mass $M_\mathrm{star} \geq 10^9 \, \Msun$ and $M_\mathrm{star} \geq 10^{10} \, \Msun$, respectively. Comparing these figures it appears clear that larger threshold masses increase the cosmic variance in the recovered signal because of the diminishing number of galaxies selected with higher mass thresholds, therefore pushing the requirements to longer $\Ltot$ to suppress cosmic variance. We have checked that this conclusion is negligibly affected by the choice of binning.

\subsection{What field of view do we need?}
Many observational surveys have limited field of view around the sightline. This can negatively impact their ability to study the \galacc, since it requires us to probe distances up to $r \sim 30$ cMpc in order to capture the extent of the flux enhancement. However, it is important to realise that such distance is not the transverse distance to the sightline ($r_\perp$), but rather the 3D distance between a pixel in the spectrum and a galaxy. Therefore, observations can probe 3D separations $r > r_\perp$, at the price of a reduced number of pixel-galaxy pairs. Therefore, here we answer the question: \textit{What is the impact of a maximum transverse distance $r_\perp^\mathrm{max}$ on the cross-correlation signal?}

\begin{figure}
    \includegraphics[width=\columnwidth]{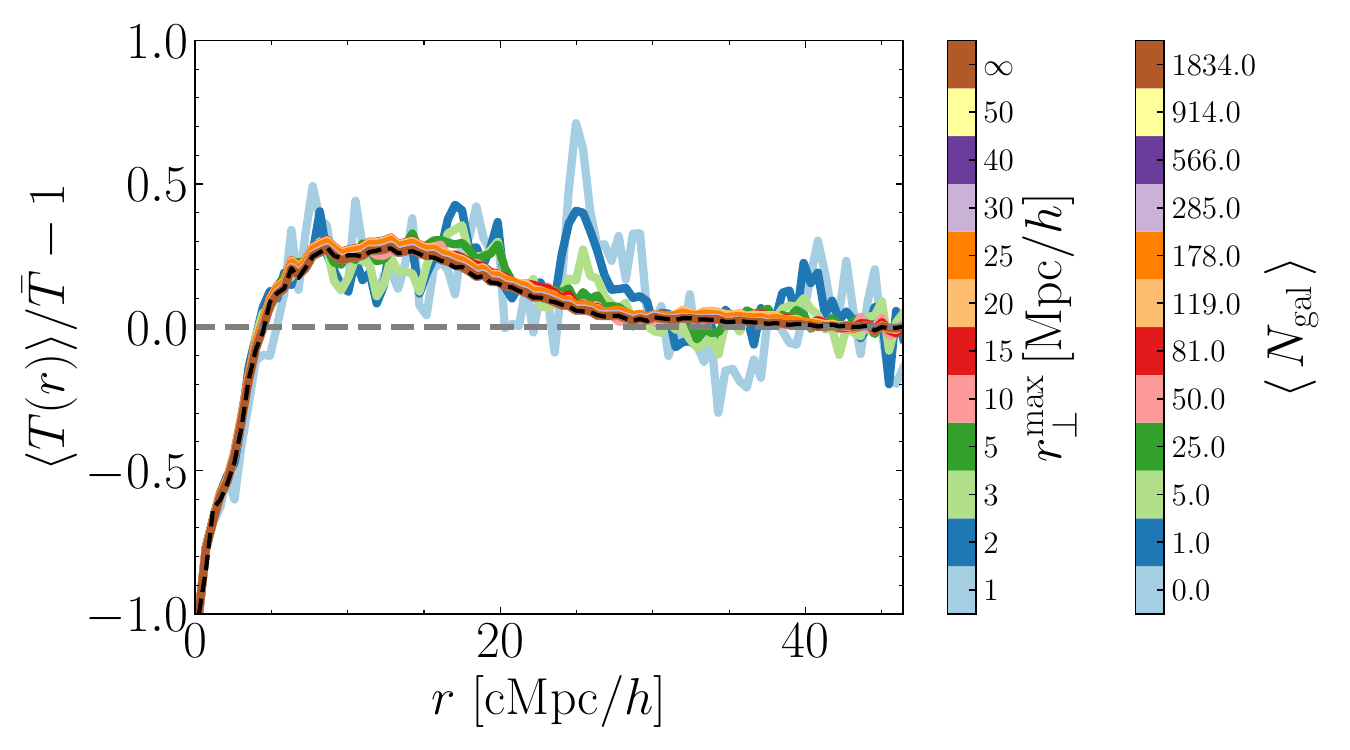}
    \caption{Impact of the maximum galaxy-sightline transverse distance ($r_\perp^\mathrm{max}$) on the prediction of the galaxy--\Lya forest cross-correlation. Different colored solid curves correspond to different values of $r_\perp^\mathrm{max}$, as reported in the inner colorbar. The black dashed line shows the result when no limit on $r_\perp$ is imposed. All curves are computed averaging together 150 spectra with total length $\Ltot = 15$ cGpc. The outer colorbar shown the average number of galaxies per spectrum within the corresponding transverse distance. We note that results are indistinguishable once $r_\perp^\mathrm{max} \gtrsim 5\,\hMpc$.}
    \label{fig:corr_vary_rt}
\end{figure}

In order to address this question, we show in Fig.~\ref{fig:corr_vary_rt} the cross-correlation obtained by imposing different values for the maximum galaxy-sightline (transverse) distance (solid colored lines). For comparison, we show the predicted cross-correlation with no maximum $r_\perp$ limit (black dashed line). It appears clear that a transverse length as small as $r_\perp^\mathrm{max} \approx 5$ is sufficient to capture the cross-correlation signal without adding any significant noise. 
However, limiting $r_\perp^\mathrm{max}$ severely decreases the number of configurations (\ie galaxy--pixel pairs) at distances $r \gtrsim r_\perp^\mathrm{max}$, therefore increasing the statistical noise. Nevertheless, our results are very encouraging for future surveys aiming at detecting this cross-correlation, as even with limited fields of view around the quasar sightline, the signal can be faithfully recovered. 
In fact, while both probing larger $r_\perp^\mathrm{max}$ and observing more sightlines increase the number of galaxy--pixel pairs, the latter allows us to probe a larger variety of cosmic environments, and it should therefore be favoured. 
Finally, limiting $r_\perp^\mathrm{max}$ also reduces the (average) number of galaxies around each spectrum used in the computation of the \galacc. We report this number for the same $r_\perp$ cuts in the outer colorbar. This, in combination with Fig.~\ref{fig:corr_vary_Llos},~\ref{fig:corr_vary_Llos_Mstar1e9},~\ref{fig:corr_vary_Llos_Mstar1e10},~\ref{fig:corr_OIII_vs_SFR_vs_Mstar} and \ref{fig:corr_gaps} (and relative discussions) gives a comprehensive overview of the impact of galaxy selection on the \galacc.

\subsection{How accurate should the spectra be?}
\label{subsec:noise_specres}
A number of surveys will soon increase the number of known high-$z$ QSO spectra. This has the potential to unlock a much broader and more systematic study of the \galacc. However, a key potentially-limiting factor is the  quality of the spectrum necessary for recovering the intrinsic signal. To determine the requirements for such observations we investigate here how the cross-correlation signal is affected by the noise and spectral resolution of the spectra. To do so, we create synthetic spectra (along exactly the same LOS used in the rest of the paper) separately including a varying degree of noise and at different spectral resolutions $R$. 
Notice that the impact of these two parameters is deeply connected. Therefore, in the following we first vary them individually in order to isolate their peculiarities, but eventually provide a joint analysis of these two, which is more useful for applications to observations. 

\begin{figure}
    \includegraphics[width=\columnwidth]{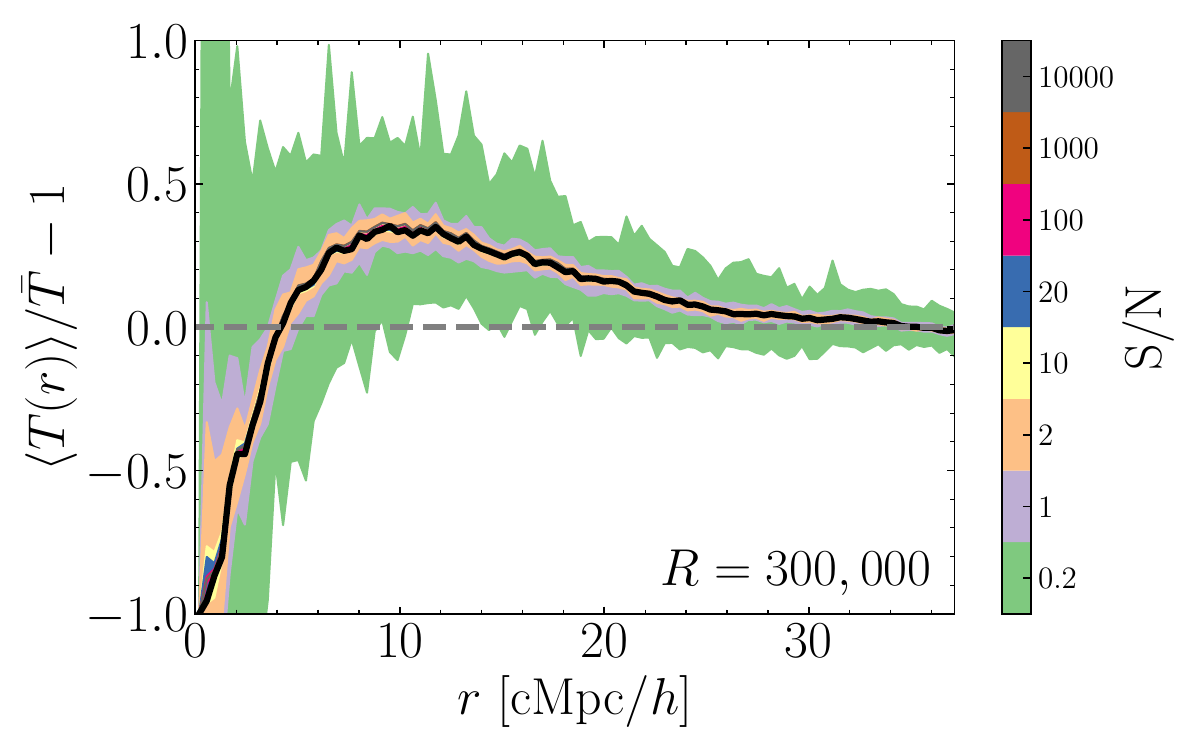}
    \caption{Impact of the spectrum S/N on the \galacc. The shaded regions show the envelope of 50 different noise realizations for each noise level. The intrinsic (\ie noise-free) signal is shown with a solid black line for reference. Even for low S/N, the intrinsic signal can be actually recovered. }
    \label{fig:corr_noise}
\end{figure}

We begin by exploring the impact of noise, assumed to be Gaussian and uncorrelated between pixels. 
The noise level is characterised by the global signal-to-noise ratio (S/N) of the continuum-normalized spectrum. In practice, we add to the normalised flux in each pixel a Gaussian noise with  $\sigma = (\mathrm{S/N})^{-1}$. We do not impose the resulting flux in each pixel to be positive (but show in Appendix~\ref{app:noise_pos_flux} the implication of doing so). Notice that, since our spectra are produced with spectral resolution of $\Delta v = 1$ km/s, the value of S/N corresponds to the average signal-to-noise ratio per resolution element. For each value of the S/N, we produce 50 different realizations of the Gaussian noise. Fig.~\ref{fig:corr_noise} shows the impact of S/N on the cross-correlation (with shaded regions reflecting the S/N). 
The figure shows that the cross-correlation signal is well recovered for S/N as low as 1 of the normalised flux. 
However, this results emerges only when showing the full envelope of possible values, while for a single noise realisation the true signal is almost-always completely lost in the noise, as discussed already. 

We note that 
in this specific case, the choice of focusing on the signal at $z=6$ has some impact. In fact, this redshift marks a sweet spot in the competition between a larger intrinsic signal \citep[achieved by moving to earlier times][]{Thesan_igm} and more numerous galaxies to beat down the statistical noise (found at later times due to structure growth). Therefore, the noise requirements will become more stringent at different times, either because of the growing impact of statistical noise or because the intrinsic signal weakens (\ie the  cross-correlation flattens on to the $T(r)/\bar{T}=1$ line).

\begin{figure}
    \includegraphics[width=\columnwidth]{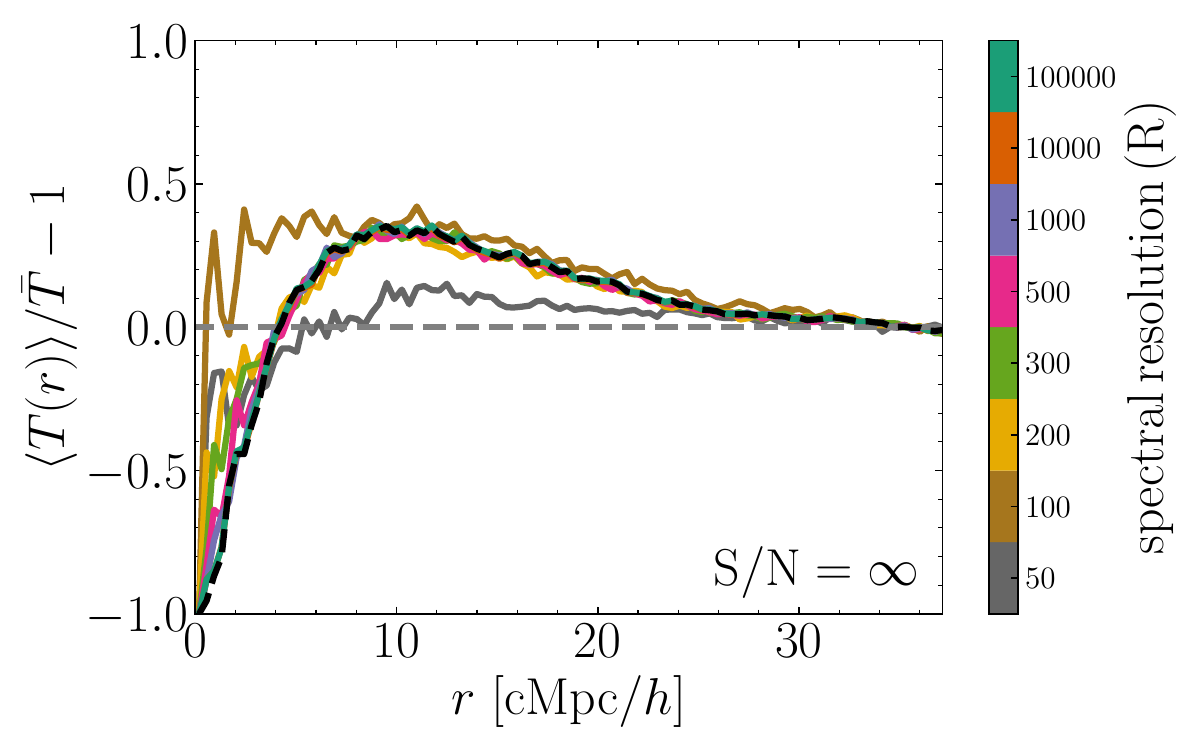}
    \caption{Impact of the spectral resolution $R \equiv \lambda / \Delta \lambda$ on the \galacc. Different lines correspond to different values of $R$, as reported in the color bar on the right. The intrinsic (\ie $R=300\,000$) signal is shown with a solid black line for reference. Even with low spectral resolution, the salient features of the cross-correlation can be recovered.}
    \label{fig:corr_R}
\end{figure}

The next property of the spectra that we investigate is the spectral resolution $R\equiv c / \Delta \varv$, where $\Delta \varv$ is the largest velocity difference that can be distinguished and $c$ is the speed of light. For each of the investigated values of $R$ we have rebinned the spectra using a boxcar filter to achieve the desired resolution (the original synthetic spectra have resolution $\Delta \varv = 1$ km/s). Fig.~\ref{fig:corr_R} shows the resulting cross-correlation for the different values, and clearly shows that a spectral resolution of $R \geq 200$ is sufficient to capture the salient features of the cross-correlation at radii $r \gtrsim 4 \, \hMpc$. A somewhat more stringent constraints is found at smaller separations ($r \lesssim 4 \, \hMpc$), where a spectral resolution of $r\gtrsim 500$ is required in order to recover the intrinsic cross-correlation. The reason is simply that at such resolution a pixel corresponds to a physical length of $\Delta l \approx 5 \, \mathrm{cMpc}/h$, and therefore smaller scales cannot be captured by the cross-correlation.

\begin{figure}
    \includegraphics[width=\columnwidth]{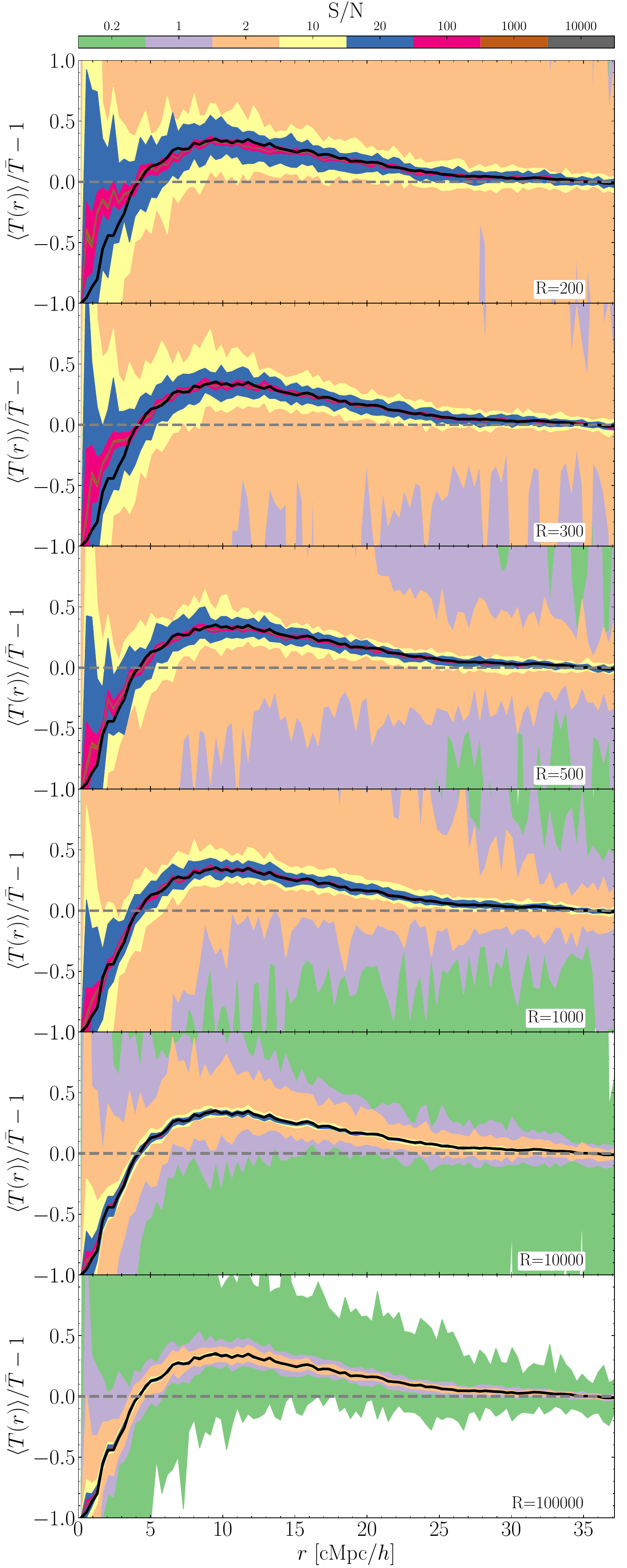}
    \caption{Joint impact of the spectrum S/N and spectral resolution $R$ on the \galacc. Each panel is a version of Fig.~\ref{fig:corr_noise} obtained for spectra of different $R$ (the same explored in Fig.~\ref{fig:corr_R}). The intrinsic (\ie $R=300000$ noise-free) signal is shown with a solid black line for reference.}
    \label{fig:corr_R_noise}
\end{figure}

The loose resolution requirements are very promising, as it renders much easier to increase the number of observed spectra. Unfortunately, the impact of resolution and noise are not independent. In fact, the resilience to noise comes from the averaging process performed while computing the cross-correlation, which averages out the impact of pixel-level noise. However, this only works as long as the regions of enhanced and suppressed transmitted flux in the spectrum contain a sufficient number of pixels. Therefore, when the spectral resolution decreases the resilience to noise worsens. To demonstrate this and to ease the task of striking a balance between the spectral resolution and S/N when attempting to recover the intrinsic signal, we show in Fig.~\ref{fig:corr_R_noise} the impact of noise spectra of different resolutions. In practice, we explore the S/N in Fig.~\ref{fig:corr_noise} (colored bands) and spectral resolutions in Fig.~\ref{fig:corr_R} (panels from top to bottom, $R$ is indicated in the bottom right of each one of them). 
As anticipated, lower $R$ are more sensitive to noise. However, even for spectral resolutions as low as $R=300$ (second panel from the top), a S/N of 20 (dark blue band) is sufficient to recover the general \galacc shape, while for $R=1000$ (third panel from the top), S/N=10 is sufficient. We remind the reader that the colored bands show the outer envelope of the \galacc computed for different noise realizations, therefore providing a worst-case scenario. In reality, most of the observed curves will be significantly closer to the noiseless one. 

In a fully-realistic observational setting, however, there are other effects that might alter -- at least quantitatively -- our conclusions. For instance, the OH airglow is a significant foreground to ground-based optical and near-infrared spectroscopic observations. Removing it might substantially increase the spectral resolution needed, in order to reduce the blending of the OH lines with the background signal.

Our results show that even with moderate noise it is possible to recover the salient features of the cross-correlation signal, shining a promising light on the possibility of significantly enlarging the number of LOS usable for this measurement. Clearly, a major issue in using sightlines for this purpose is the need to identify galaxies around them. Overall, these results, combined with the one provided in Sec.~\ref{subsec:how_many}, outline the observational requirements for the reliable characterization of the \galacc. In Sec.~\ref{sec:future} we will discuss in more details which surveys (available or forthcoming) can be used for this measurement. 

\subsection{How to select galaxies?}
\label{subsec:corr_OIII_vs_SFR_vs_Mstar}
Observations of the \galacc employ different techniques to identify galaxies, ranging from \CIV absorption in the same spectrum to \OIII emission around the sightline. 
Therefore, it is unclear how to properly compare such observations among themselves and with simulations. 
For this reason, here we have decided to remain agnostic and select galaxies based on their stellar mass. However, it is key to determine whether these different selection techniques are biasing  the inferred \galacc . Additionally, it is foreseeable that in the coming years more and more surveys will identify galaxies through their emission lines. The \OIII emission line is particularly promising, and is in fact used by a number of ongoing JWST surveys. For this reason, we study here the effect of selecting galaxies based on their star formation rate (SFR), stellar mass ($M_\mathrm{star}$) and \OIII flux. For the latter, we employ the synthetic galaxy SEDs described in \citet{Thesan_lim, Thesan_data}. Thanks to these, we can restrict the galaxy selection to the brightest \OIII emitters in the simulation box, defined here based on their equivalent width (EW$_\mathrm{OIII}$). 

\begin{figure}
    \includegraphics[width=\columnwidth]{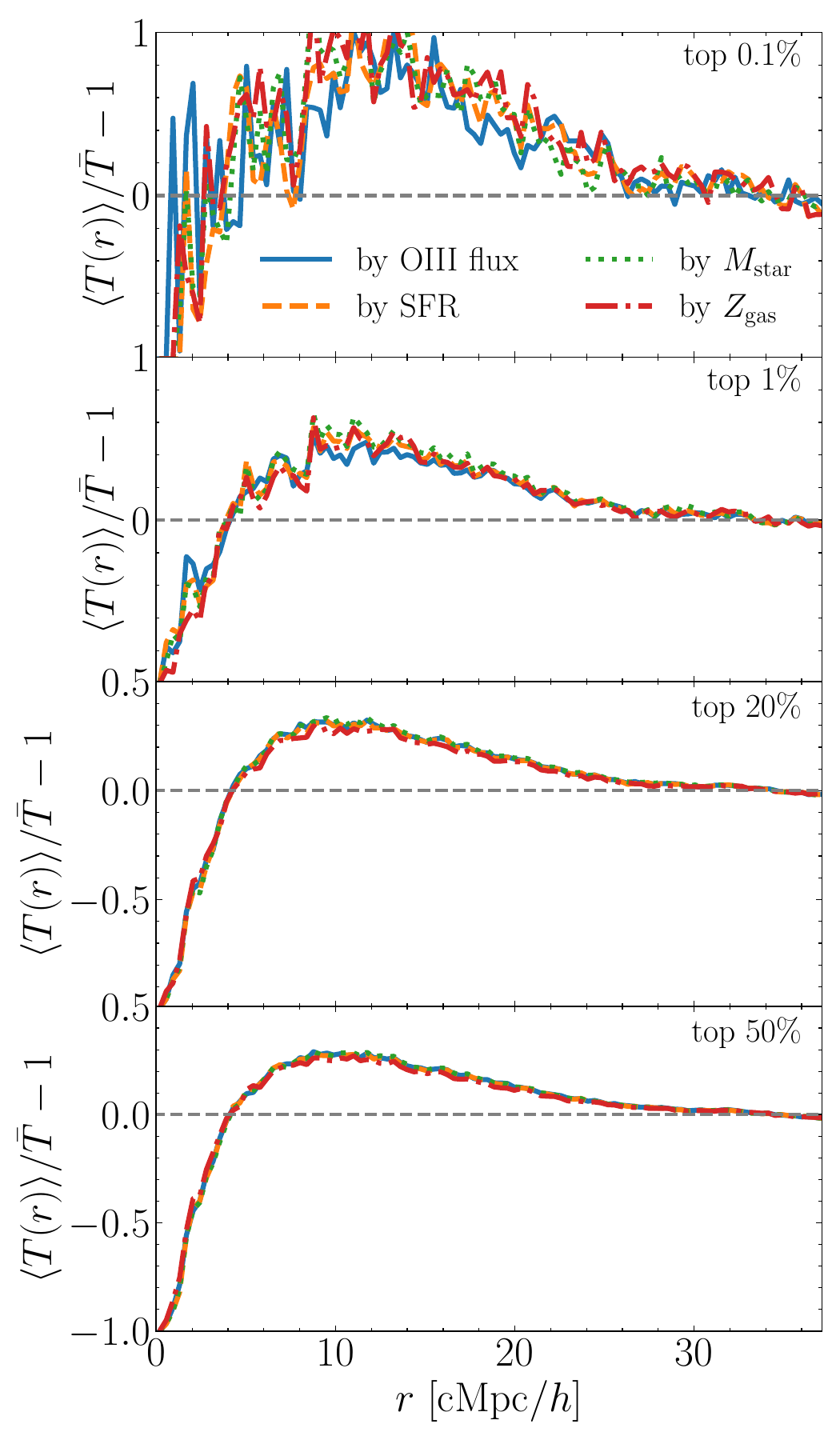}
    \caption{\galacc computed for the top 0.1\%, 1\%, 20\% and 50\% (top to bottom panels, respectively) galaxies in \thesanone, selected on the basis of their \OIII flux in the synthetic SED (solid blue line), SFR (dashed orange line), stellar mass (green dotted line) and gas-phase metallicity (red dot-dashed line). The different selection methods do not impact the recovered signal.}
    \label{fig:corr_OIII_vs_SFR_vs_Mstar}
\end{figure}

While the limited volume of our simulations hinders a full one-to-one comparison with \eg the EIGER survey, it still provides precious information on the biases carried by such selection. 
In Fig.~\ref{fig:corr_OIII_vs_SFR_vs_Mstar} we show the cross-correlation computed selecting the top 0.1\%, 1\%, 20\% and 50\% (panels from top to bottom) galaxies when ranked in \OIII flux, SFR, stellar mass and gas metallicity (as a simple proxy for the strength of their \CIV absorption signature in the QSO spectrum). Notice that the galaxy SED were computed only for a subset of `well-resolved' galaxies \citep[as defined in][]{Thesan_lim}, resulting in approximately 10000 galaxies at $z=6$. Therefore, to enable a proper comparison, we have restricted \textit{all} selection to the same subset of galaxies, despite the SFR, stellar mass and gas metallicity being available for all simulated galaxies. 
The different panels show a consistent picture, namely that the selection performed does not affect the resulting cross-correlation. More selective criteria result in a more noisy signals (because of the smaller number of objects resulting in larger noise) and higher peaks \citep[because the sources are more biased and more luminous, as discussed in Sec.~\ref{subsec:delta}][]{Thesan_igm}, but there is no discernible difference between selecting galaxies based on their stellar mass, star formation rate or \OIII flux. 

The insensitivity of the \galacc to the source selection stems from two conspiring reasons. First, in the \thesan simulation most galaxies lie on the star-formation main sequence and on the mass-metallicity relation \citep{Thesan_data}, with only very few deviating from it \citep{Thesan_sizes}. Therefore, the most star forming galaxies are also the most massive ones (in terms of stellar component) and the most metal-rich ones. Second, the large volume covered by the simulations does not allow to resolve the ISM of the simulated galaxies. Therefore, the line emission has to be `painted' on the unresolved ISM using approximate methods \citep{Thesan_lim}. This entails that, for the \OIII line, there is a built-in correlation between its flux and the star formation of the galaxy. While the complex radiation transport and dust obscuration entering in the SED production can break this correlation, it is the case that the \OIII flux and the galaxy SFR are tightly related in \thesan \citep{Thesan_lim}. These two effects conspire to essentially yield the same galaxy sample, regardless of the selection method. In fact, the overlap between the samples of galaxies resulting from the three selection methods discussed is larger than 85\% in all but one case. We caution, however, that this result might change if we were able to resolve the ISM of galaxies in a large-enough volume to compute the \galacc. Unfortunately this is beyond the reach of current simulations. 

The results presented above strengthen the approach taken in this paper and in previous theoretical analysis, \ie selecting galaxies based on their stellar mass. It also provides precious information for future observational studies.

\subsection{Impact of gaps in the galaxy catalog and spectrum}
\begin{figure}
    \includegraphics[width=\columnwidth]{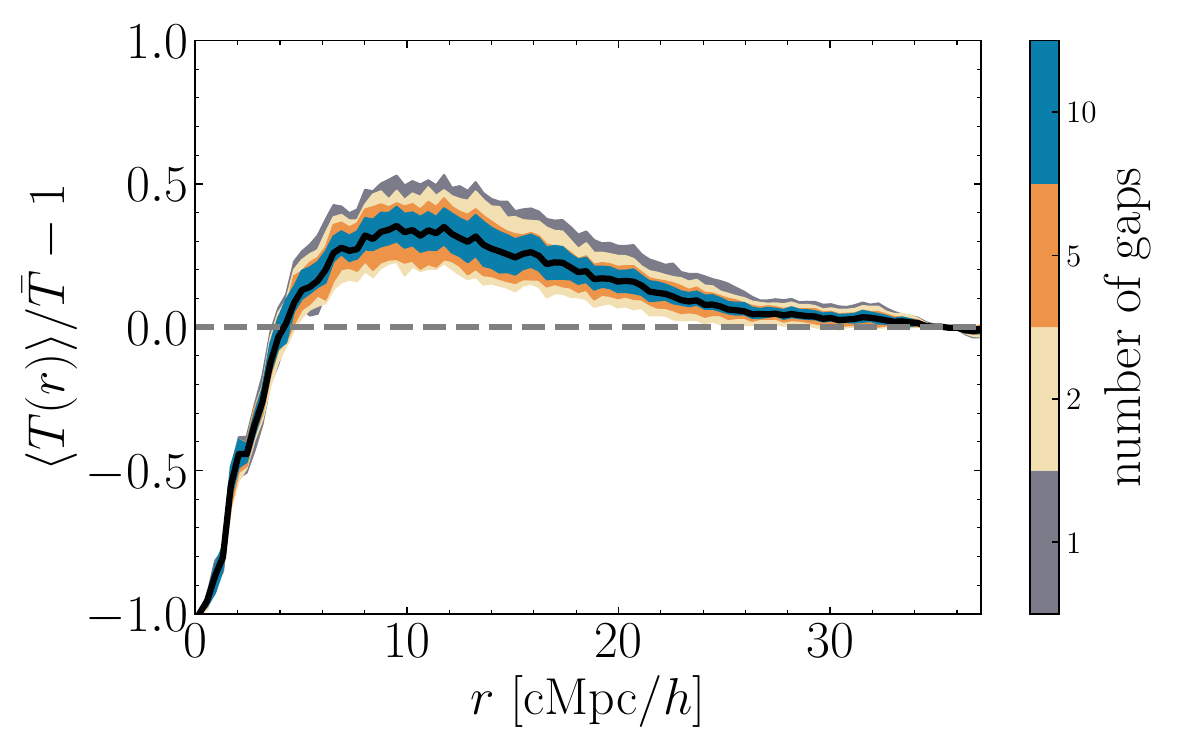}
    \includegraphics[width=\columnwidth]{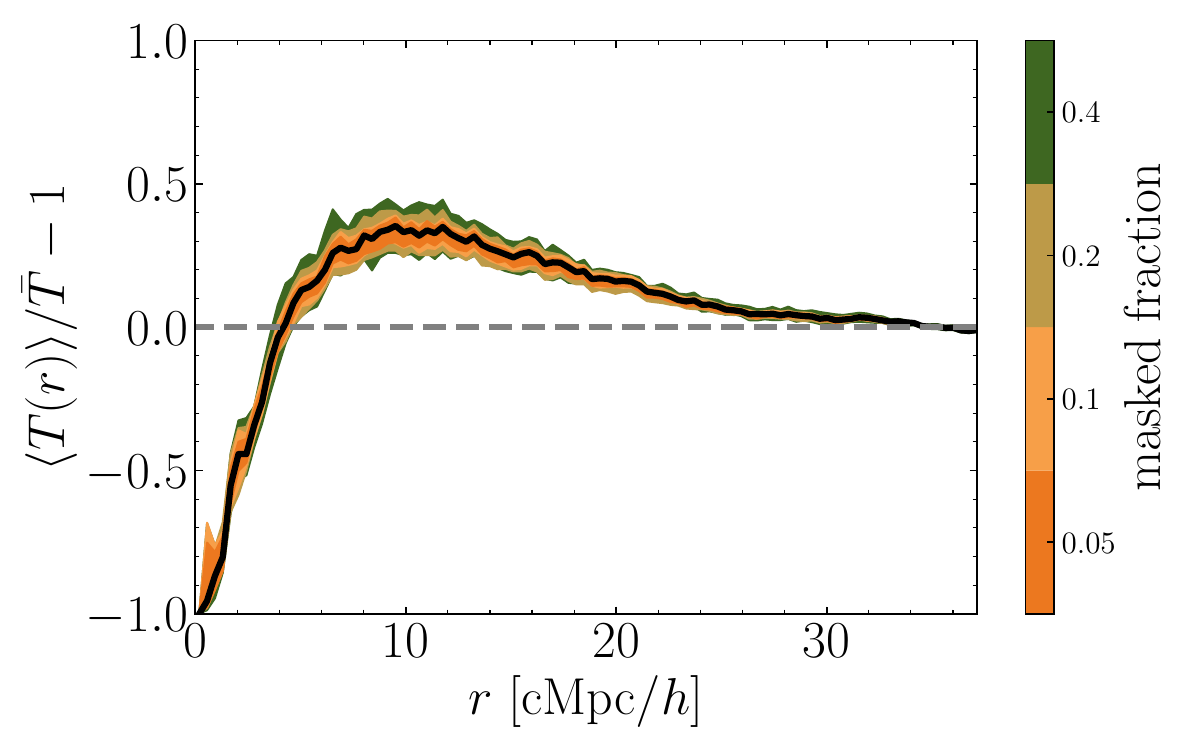}
    \caption{Impact of changing the number of gaps in the galaxy distribution around each line of sight (top) or the total masked fraction in the observed spectrum (bottom). The solid black line shows our fiducial signal (computed from 300 lines of sight without masking out any galaxy), while coloured regions report the outer envelope of the \galacc computed from 50 random realizations of gaps, with different colors referring to different number of gaps (and therefore average gap length, see text). Fewer longer gaps in the galaxy distribution are more impactful than numerous shorter ones, while gaps in the spectrum appear to be hardly significant. }
    \label{fig:corr_gaps}
\end{figure}

A potentially problematic case that can arise in observations of the \galacc is when the galaxy detection method employed results in gaps in the galaxies identified around the line of sight. This can arise \eg if a set of different emission lines is employed to identify galaxies at different redshifts (\ie at different positions along the line of sight) because of the finite spectral coverage of the instrument employed. We test the impact of such an artefact in the galaxy selection by manually removing galaxies in specific (small) redshift windows along a sightline. Specifically, we sample the length of each such gap from a uniform distribution in the range $[0, 0.5 \times L_\mathrm{spec}/N_\mathrm{gaps}]$, where $L_\mathrm{spec}$ is the length of each spectrum and $N_\mathrm{gaps}$ is the number of gaps in each spectrum. This ensures that, on average, $25 \%$ of the galaxies are masked out. We produce 50 random realizations of gaps and show in the top panel of Fig.~\ref{fig:corr_gaps} the envelope of all resulting \galacc for varying number of gaps. We find that fewer (and hence longer) gaps more strongly impact the inferred \galacc than more shorter ones. 
We also test the impact of fixing the gap length and changing the gap number. This shows results fully consistent with the changing number of galaxies selected (since in this case the average masked-out region increases with gap number). 

We also test the case in which it is instead the spectrum to have gaps, due for example to removal of foreground contamination. Since in this case it is more likely to have a large number of small gaps rather than one or few very large ones, we instead quantify the impact of masking different fractions of the spectrum, sampling the gaps length from the range  $[0, 0.05 \times L_\mathrm{spec}]$ (note that, unlike in the galaxy gaps case, this does not depends on the gap number). We show in the bottom panel of Fig.~\ref{fig:corr_gaps} that the \galacc is very resilient to this type of masking, likely because the number of pixels in the spectrum is sufficiently high so that, even when a large fraction of them is removed, the underlying signal can be recovered with small uncertainty. We have also checked that, as expected, performing the same gap number-dependent sampling of the gap lengths as we have done for the galaxy catalog results in almost identical results.

\subsection{Lightcone effects}
\label{subsec:lightcone_effects}
\begin{figure}
    \includegraphics[width=\columnwidth]{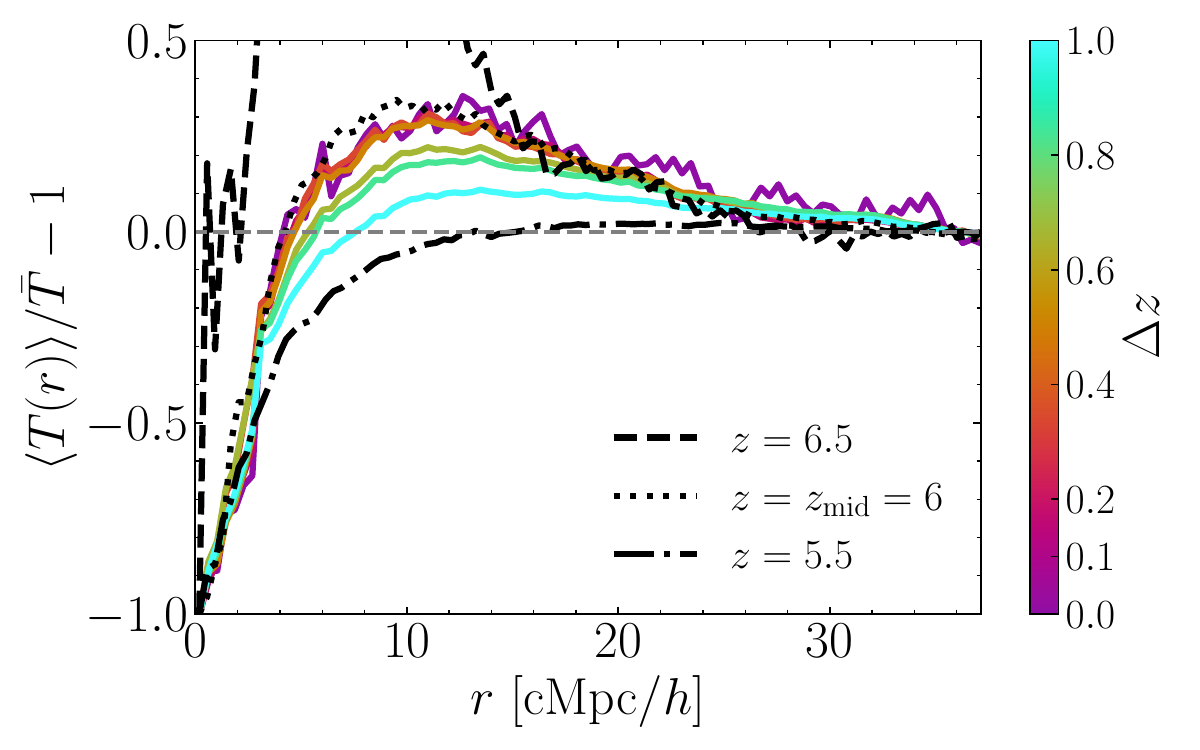}
    \caption{Impact of the width of the redshift window employed for the computation of the \galacc (from lightcone-like lines of sight). The window is defined as $[z_\mathrm{mid} - \Delta z /2, z_\mathrm{mid} + \Delta z /2)$, where for this figure we employ $z_\mathrm{mid} = 6$ and the different values of $\Delta z$ are reflected in the different color of the lines, as shown by the color bar on the right-hand side. The dashed, dotted and dot-dashed black lines show the cross-correlation signal at fixed redshift (\ie without lightcone effects) at $z=6.5, 6, 5.5$ respectively.}
    \label{fig:corr_lightcone_effects}
\end{figure}

In order to test the impact of lightcone effects on the \galacc, we have included them in our synthetic \Lya forest spectra and galaxy catalog (using a step-wise constant approximation). This allows us to explore the impact of the redshift window $[z_\mathrm{mid} - \Delta z/2, z_\mathrm{mid} + \Delta z/2]$ employed in observation of this quantity. We show in Fig.~\ref{fig:corr_lightcone_effects} the resulting cross-correlation for a redshift window centred at $z_\mathrm{mid}=6$ and of varying width $\Delta z$ (solid colored lines). For reference, we also show the cross-correlation computed using our coeval (i.e. without any lightcone effect) at $z=6.5, 6, 5.5$ (dashed, dotted and dot-dashed black lines, respectively). Note that, since for these coeval sightlines the entire synthetic spectrum lies at a fixed redshift, the total path length covered by these spectra is larger (for the same number of sightlines considered, as it is the case here). Therefore, we expect small differences in the cross-correlation computed from the lightcone spectra fixing $\Delta z \approx 0$\footnote{Since for $\Delta z = 0$ the spectrum length is vanishing, we have approximated this by considering only the part of each spectrum associated to $z_\mathrm{mid}$, which is non-vanishing because of the stepwise-constant approximation used in producing the lightcone-like spectra (see Sec.~\ref{subsec:spectra}). In practice, thanks to our high time cadence of snapshots, this is equivalent to $\Delta z \approx 0.05$.} and the one computed from the coeval spectra. However, if we were to match the spectrum path (in both length and position in the simulation box), the result from the lightcone spectra would be \textit{by construction} identical to the one from the coeval sightlines.

We have chosen to focus on $z_\mathrm{mid} = 6$ for a number of reasons. First, observations of the \galacc \citep[\eg][]{Kakiichi+2018,Meyer+2019,Meyer+2020,EIGERI} are at or close to this redshift. Second, it approximately marks the beginning of the tail end of reionization, characterised by a rapid evolution of the volume-averaged \HI fraction as the ionized regions overlap. This makes such period especially interesting for a number of reasons: (\textit{i}) it induces a rapid evolution of the IGM global quantities, enabling an accurate timing of the EoR and, in turn, tight constraints on the source properties; (\textit{ii}) the flux in the \Lya forest is non-negligible (unlike at \eg $z\gtrsim7$) but the fluctuations in the UV field are still strong enough to leave imprints in the galaxy--\Lya correlation \citep[unlike at \eg $z\lesssim 5$][]{Meyer+2020}. Third, since our simulations end at $z=5.5$, choosing $z_\mathrm{mid} = 6$ allow us to study the impact of redshift windows up to $\Delta z = 1$ without running into the issue that no more outputs are available. 

Fig.~\ref{fig:corr_lightcone_effects} shows that the impact of lightcone effects is negligible for $\Delta z \lesssim 0.4$, essentially only affecting the smoothness of the curve thanks to improved statistics when allowing for larger redshift bins, that therefore include more galaxies. However, for larger $\Delta z$ the lightcone effects become relevant. In particular, the amplitude of the excess is reduced and the position of the peak moves to larger distances. This occurs because the signal becomes dominated by the lower-redshift part of the redshift window as a consequence of the larger number of galaxies (and therefore pixel-galaxy pairs) found in it. In fact, the curve approaches more and more the dot-dashed lines showing the cross-correlation signal recorded at the lowest redshift encapsulated in the largest redshift window. 

\begin{figure}
    \includegraphics[width=\columnwidth]{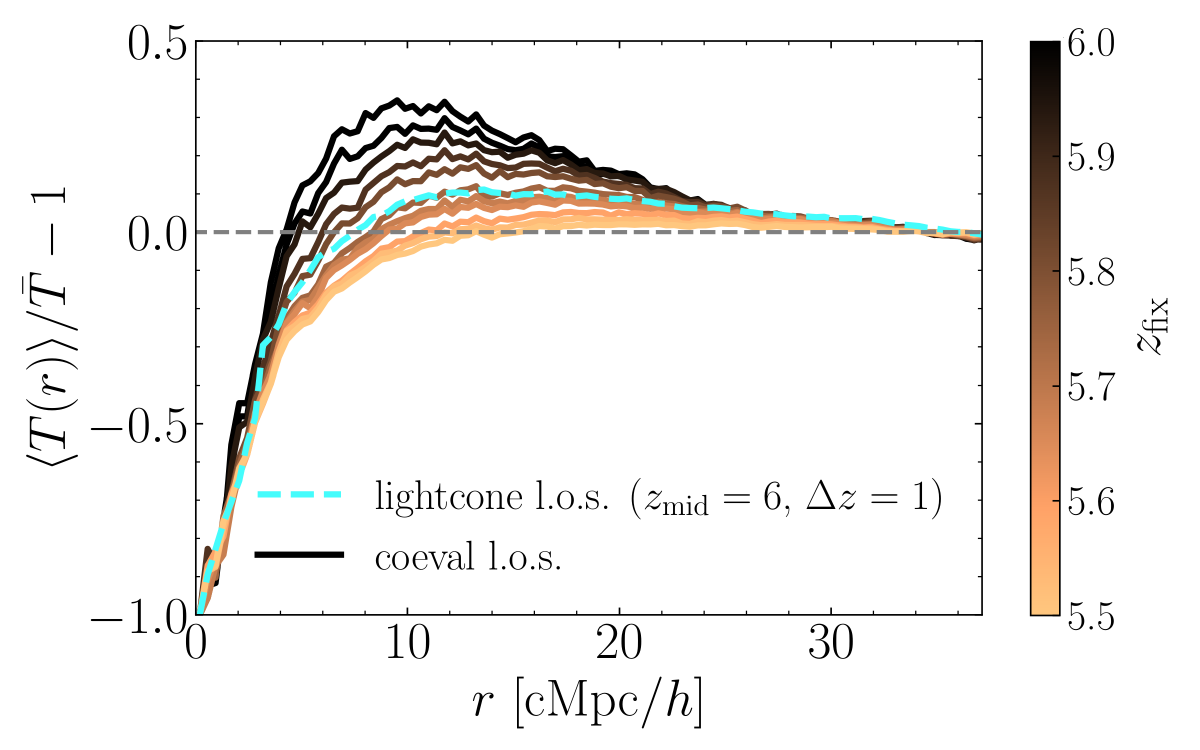}
    \caption{Comparison between the galaxy--\Lya cross computed in the redshift window $[5.5, 6.5)$ (including lightcone effects, blue dashed line) and the same quantity computed for different fixed redshifts $z_\mathrm{fix}$ (solid lines with colour varying to reflect their redshift). The lightcone-like signal is very close to the one obtained at some effective redshift, that differs from the central one of the redshfit window ($z \approx 5.7$ in this case).}    \label{fig:corr_lightcone_effects_vs_fixz}
\end{figure}

This has important implications for studies that cover broad redshift windows, \eg \citet{Meyer+2019,Meyer+2020}. In particular, it implies that in order to properly compare such observations with theoretical and numerical predictions, lightcone-like sightlines are needed. This requirement was not met by previous studies \citep[\eg][]{Thesan_igm}, but our results demonstrate how this is necessary for an accurate comparison. 
Alternatively, an approximate workaround is to compare observation to prediction made at an effective redshift $z_\mathrm{eff}$ encapsulating the biased mixing of signals coming from different redshift ranges. In order to investigate this possibility, we show in Fig.~\ref{fig:corr_lightcone_effects_vs_fixz} the cross-correlation signal computed from the lightcone LOS using $\Delta z = 1$ (dashed blue curve, same as in Fig.~\ref{fig:corr_lightcone_effects}) alongside the correlation signal computed at fixed time (solid lines) for a variety of redshift (indicated by the colorbar). It can be seen that the cross-correlation computed from coeval sightlines at $z=5.73$ almost perfectly coincides with the one on obtained from the lightcone LOS using the redshift window $z \in [6.5, 5.5]$. From the previous discussion, it follows that a good candidate for this  $z_\mathrm{eff}$ is the median redshift of the galaxies used to compute the cross correlation. We have tested that this is indeed the case even for the largest values of $\Delta z$ probed ($\Delta z =1$, see also Appendix~\ref{app:zeff_zmed_zmid} for more details). 

Finally, we note that in \citet{Thesan_igm} the predictions of the \thesan simulations were compared the data collected by \citet{Meyer+2019}, that compound observations over the broad redshift window $4.5 < z < 6.3$. In that study, the simulations output at $z=5.5$ were used, as this is very close to the central redshift of the observed data. In light of the results discussed here, it would have been more appropriate to compare to a lower redshift, that however is unfortunately not available for the \thesan simulations. Nevertheless, extrapolating the redshift evolution found in \thesan \citep[Fig.~16 of][]{Thesan_igm}, we can speculate that using a lower-redshift output for the comparison would worsen the tension found between the simulated and observed cross-correlation, requiring an even later end of reionization to match the observed signal.

\section{Physical processes connected to the \galacc}
\label{sec:results_phys}

\begin{figure}   
    \includegraphics[width=\columnwidth]{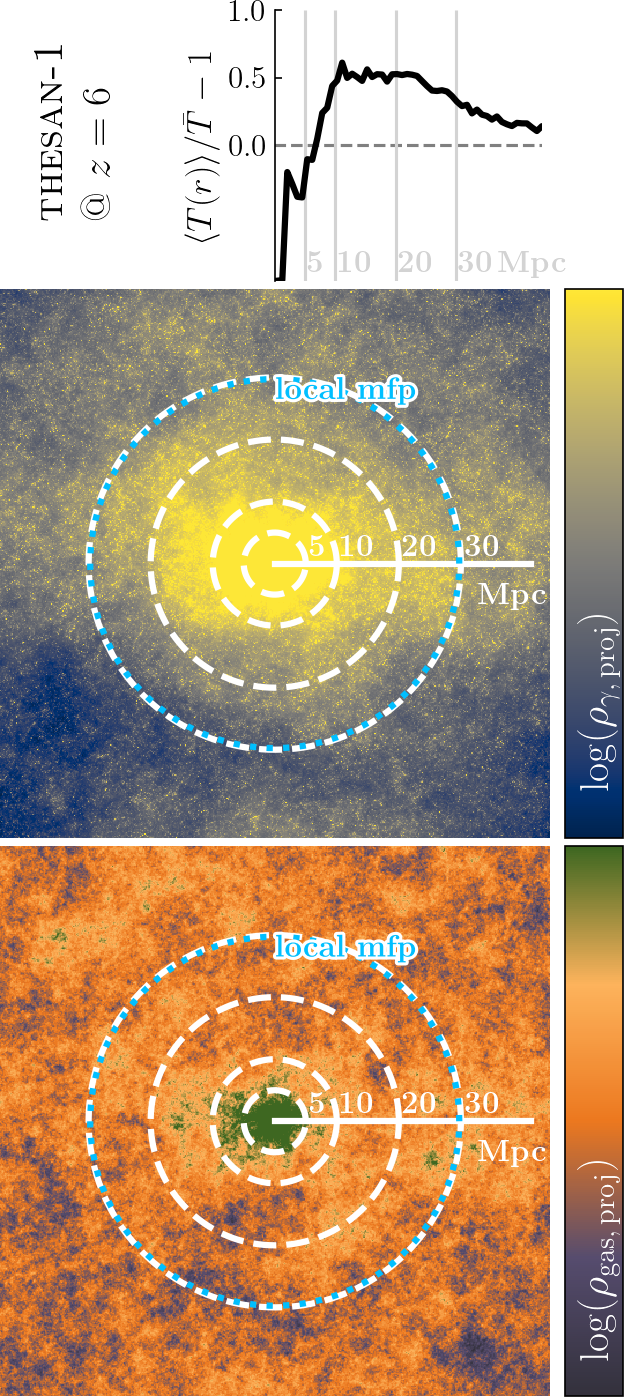}
    \caption{Galaxy--\Lya cross-correlation computed from \thesanone at $z=6$ (top panel) using all galaxies with stellar mass $M_\mathrm{star} \geq 10^9 \, \Msun$. We have marked four radial scales using vertical grey lines. In the middle and bottom panel, respectively, we show the stacked 
    ionizing photons density 
    and gas density around the same galaxies used to compute the \galacc, projected along one axis for visualization purposes. The colorbar depend logarithmically on the projected quantity value. Dashed white circles mark the same scales as above, while the blue circle shows the \textit{local} (\ie around the stacked haloes) mean free path of ionising photons. The inner circle nicely marks the scales where the galaxy overdensity (bottom panel) triggers a sharp drop in the transmitted flux, due to the enhanced recombination rate. The outer circle, instead, corresponds to the approximate scale where the enhancement ionization (top panel) due to the proximity to the sources of ionizing photons powers an enhanced Lyman-$\alpha$ transmission. 
    }
    \label{fig:radial_stacks}
\end{figure}

The leading interpretation of the \galacc paints a picture where the observed signal arises from two opposite effects, each dominating at different scales around galaxies. The enhanced ionizing photon density provided by (large) galaxies boosts the transmission of \Lya in their surroundings. This transverse proximity effect can be detected statistically by looking at the excess \Lya transmission around galaxies. 
At the same time, the overdensity where galaxies reside suppress the \Lya transmission, thanks to their increased density and, hence, boosted \HII recombination rate. This is the dominant effect at the smallest scales, while the aforementioned radiation proximity effect takes over at intermediate ones. 
In Fig.~\ref{fig:radial_stacks} we visually show this by reporting in the top panel the \galacc computed from \thesanone at $z=6$ and using galaxies with stellar mass $M_\mathrm{star} \geq 10^9 \, \Msun$ (black solid line). We mark the radii corresponding to $5$, $10$, $20$, and $30$ Mpc with light grey vertical solid lines. In the middle and bottom panels, we show the stacked ionizing photon density and gas density around the same galaxies used in the computation of the \galacc. In order to compute these quantities, we extract gas properties in a cubic region with side $45$ Mpc and centered on each galaxy in the sample. We then stack these regions and finally project the resulting cube along one axis for visualisation purposes. In the middle and bottom panel, we mark with dashed circles the same radii as above. It can clearly be seen that the central regions (approximately within $5$-$10$ Mpc) are dominated by high densities, which suppress the \Lya transmission thanks to the boosted recombination rate suppressing the ionized fraction. At intermediate scales ($10 \lesssim d / \mathrm{Mpc} \lesssim 30$) the density has already fallen to the background level, but the ionizing photons density is still largely above the one observed farther away from the center. The latter enhances the \Lya transmission at such scales, resulting in a broad peak in the \galacc.

\subsection{Connection to the local overdensity}
\label{subsec:delta}
As discussed in Sec.~\ref{sec:results_obs}, the selection of galaxies to observe has an impact on the resulting cross-correlation signal. Here we aim to address the questions: \textit{What do we learn by looking at different galaxy populations? Is there an optimal galaxy sample?} In \citet{Thesan_igm} it was already shown that the amplitude of the flux enhancement depends on the galaxy used to compute the cross-correlation. In particular, more biased tracers (\ie more massive galaxies and haloes) yield a stronger signal. At the same time, it was also shown that, once the reionization history is accounted for, models where small galaxies dominate the photon budget produce approximately the same \galacc as models dominated by large galaxies. To elucidate such a counter-intuitive result, here we supplement that study with by investigating the impact of the local overdensity of the galaxies selected. In fact, a number of relevant physical processes relevant for the cross-correlation signal investigated here are sensitive to the local overdensity. First, higher local densities imply larger recombination rates that can suppress the flux in the \Lya forest. Second, overdense regions hosts more galaxies, whose radiation output combines to create larger ionized bubbles and stronger radiation fields. Finally, larger overdensities are more likely to host massive galaxies. 

\begin{figure}
    \includegraphics[width=\columnwidth]{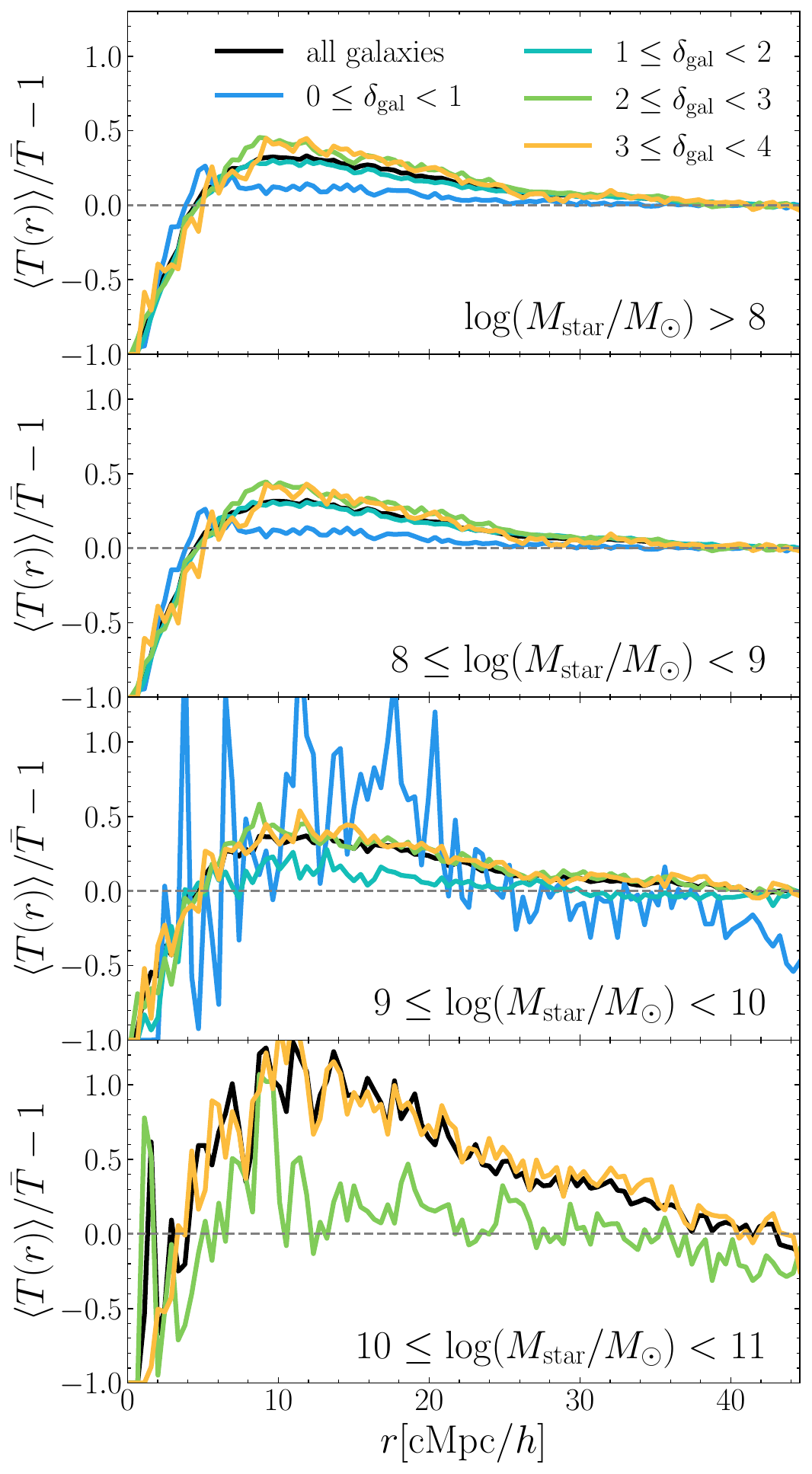}
    \caption{\galacc computed using galaxies of different stellar masses (panels from top to bottom, as reported in their top right corner) and residing in different overdensities ($\delta_\mathrm{gal}$, solid lines of different colors). Galaxies in overdense regions are the main source of the flux modulation at intermediate distances, thanks to their concerted effort significantly ionizing the surrounding medium over the Universe average, which can not be (as effectively) achieved by individual objects.}
    \label{fig:corr_delta}
\end{figure}

For the purpose of this study, we define the local overdensity of a galaxy ($\delta_\mathrm{gal}$) as in \citet{Thesan_bubbles}. In practice, we first smooth the density field with a Gaussian filter with standard deviation 1 cMpc. We then define $\delta_\mathrm{gal} \equiv (\rho_\mathrm{smooth} - \bar{\rho}) / \bar{\rho}$, where $\bar{\rho}$ is the average gas density in the Universe at the given redshift and $\rho_\mathrm{smooth}$ is the smoothed density field. Equipped with this definition, we compute the cross-correlation signal in bins of stellar mass and local overdensity. More specifically, the former are delimited by $M_\mathrm{star}/M_\odot = 10^8, 10^9, 10^{10}, 10^{11}$ and the latter by $\delta_\mathrm{gal} = -1, 0, 1, 2, 3, 4, 5$. We show the resulting correlation function in Fig.~\ref{fig:corr_delta}, where the different panels correspond to different bins in stellar mass (reported in the bottom right corner of each panel). The lines within each panel show the result of isolating different local overdensities (shades of green) and of considering all galaxies together (black line). The effect of the local overdensity can be quite strong. Larger $\delta_\mathrm{gal}$ increases the \textit{amplitude} of the measured \galacc, but does not affect its shape. 

The finding described above has important implications for the interpretation of the observed signal. In fact, a random sightline to a QSO will cross (or pass in the proximity of) regions of different overdensity, each one imprinting its own intrinsically different signature, contributing to the `structure formation noise' described in Sec.~\ref{subsec:how_many}. This is not problematic if the cross-correlation is computed over a sufficiently-large number of sightlines, but can bias the results if this number is limited. In this case, the impact of a sightlines sampling an extreme overdensity can be very large, which renders it difficult to connect the observed signal to the physical properties of the Universe. Similarly, if the galaxy detection method employed results in galaxies spatially clustered around few specific positions of the spectrum (\eg when observations are sensitive to just extremely bright objects, which tend to be clumped in large overdensities), the final result will be driven by one or few overdensitied where such objects reside.

\subsubsection{Insensitivity to source bias}
The results described above are linked to an intriguing finding of \citet{Thesan_igm}, namely that once the different reionization histories are factored out, the cross-correlation signal does not depend on the sources of reionization. In fact, in their Fig.~17, the signal from the \thesanlow and \thesanhigh simulations (which emit ionizing photons only from galaxies more and less massive than $10^{10} \, \Msun$, respectively) are essentially indistinguishable from the one stemming from their standard runs. This appears at odds with their result that galaxies produce a modulation of the \Lya transmission in their local environment that scales with their mass. These two apparently-contradicting statements can be reconciled once the role of overdensity is taken into account. For example, in the case of \thesanlow, although only small galaxies are emitting ionizing photons, they are still preferentially found in overdensities, where also larger objects tend to reside. This, combined with their larger photon output (\ie higher escape fraction, needed to achieve a somewhat similar reionization history) entails that overall the contribution to the \galacc still remains dominated by large conglomerates of small galaxies, that essentially act as rare bright sources in this model.

\begin{figure}
    \includegraphics[width=\columnwidth]{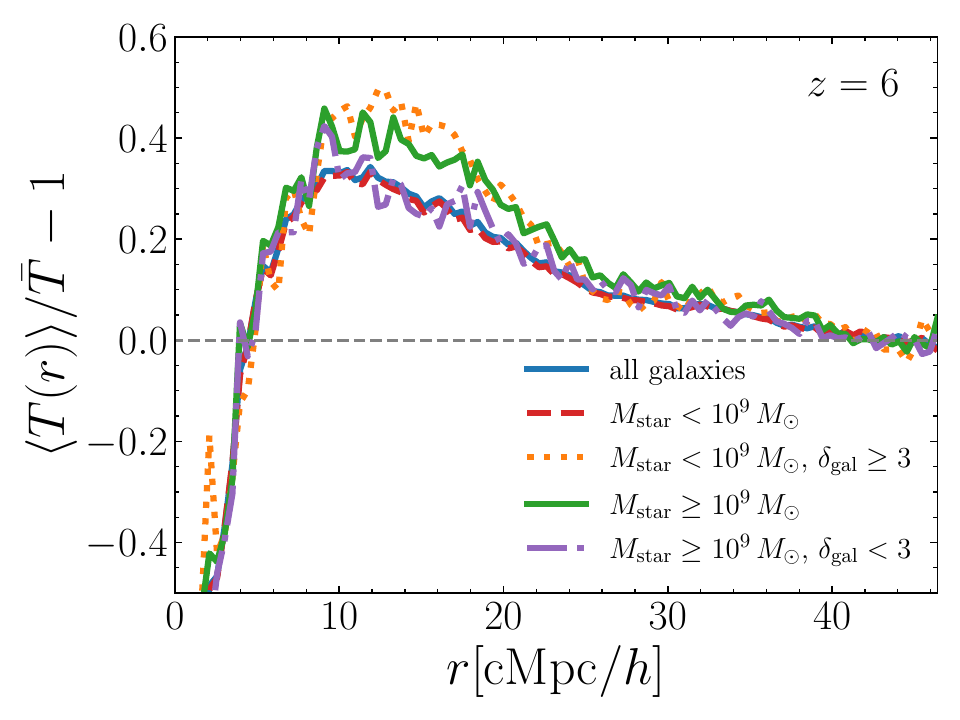}
    \caption{Direct comparison of the impact of galaxy stellar mass and overdensity on the \galacc. 
    Different lines show different selection of galaxies based on their stellar mass and local overdensity. 
    }
    \label{fig:corr_delta_small_big_galaxies}
\end{figure}

In order to show this explicitly, in Fig.~\ref{fig:corr_delta_small_big_galaxies} we display the cross-correlation signal computed at $z=6$ for all galaxies (solid blue line), for small galaxies (\ie $M_\mathrm{star} < 10^9 \, \Msun$, dashed red line), for small galaxies with local overdensity $\delta_\mathrm{gal} \geq 3$ (dotted yellow line), for large galaxies (\ie $M_\mathrm{star} \geq 10^9 \, \Msun$, solid green line) and for large galaxies with local overdensity $\delta_\mathrm{gal} < 3$ (dot-dashed purple line). A first striking feature is that the exclusion of big galaxies does not affect the correlation signal at all (compare the solid blue and dashed red lines), clarifying that the impact of large objects in term of ionizing photons is negligible \citep[as expected from the ionizing photons budget discussed in][]{Thesan_intro}. However, selecting large galaxies to compute the cross-correlation \citep[as done in][]{Thesan_igm} means selecting preferentially overdense regions (because of the larger bias associated to such objects). The impact of such selection can be seen comparing the dashed red and dotted yellow line. Both curves are computed only for small galaxies, but the latter additionally selects only objects with $\delta_\mathrm{gal} \geq 3$, resulting in a significantly boosted signal. In fact, such boost is very similar to the one obtained by selecting large galaxies only (solid green line). Symmetrically, removing galaxies in large overdensities from the sample of massive objects (dot-dashed purple line, \ie artificiallty removing the preference of large galaxies to reside in overdense regions) brings the signal down to a level very similar to the one obtained including all galaxies. 

From Fig.~\ref{fig:corr_delta_small_big_galaxies} and the related discussion above, it appears that the main driver of the transmissivity enhancement at intermediate scales are overdense regions and not bright sources. Naturally, these two are related in the \LCDM cosmology, since larger/brighter sources are more biased than smaller/fainter ones and preferentially reside in overdense regions. However, our results show that it is the clustering of multiple faint sources (which occurs primarily in overdense regions) to produce the observed signal, not individual/rare bright sources, as a consequence of the complexity of the \Lya signal and of the additional recombinations occurring in overdense regions counteracting the additional ionizing photons emitted by the rare bright sources \citep[which are sub-dominant with respect to the one emitted by smaller galaxies, see \eg][]{SPHINX20,Kostyuk+23,Thesan_fesc}. This finding reconciles the apparently contradicting results that the ionizing photons budget is dominated by small galaxies, that the \galacc is enhanced around bright objects and that restricting the ionizing photons production to the largest/smallest galaxies does not affect the signal (once the different reionization histories are factored out).

\subsection{Dependence on the sightline effective optical depth}
The IGM ionized fraction along the line of sight is expected to affect the inferred \galacc. In fact, more neutral sightlines are more opaque to \Lya photons (\ie their effective optical depth $\taueff$ is larger), decreasing the total transmitted flux. For this reason, following \citet{Garaldi+2019croc} the \galacc is typically normalised by the average transmitted flux, in order to factor out sightline-to-sightline variations in the average density (and, hence, neutral hydrogen fraction). However, the choice of whether to compute the average transmitted flux from the observed spectra sample or to use some independent measurement can create artifacts that need to be understood. We stress that, as outlined in Sec.~\ref{subsec:computing_galacc}, this distinction is practically non-existent for our work, by virtue of how the number and production method of our synthetic spectra.

\begin{figure}
    \includegraphics[width=\columnwidth]{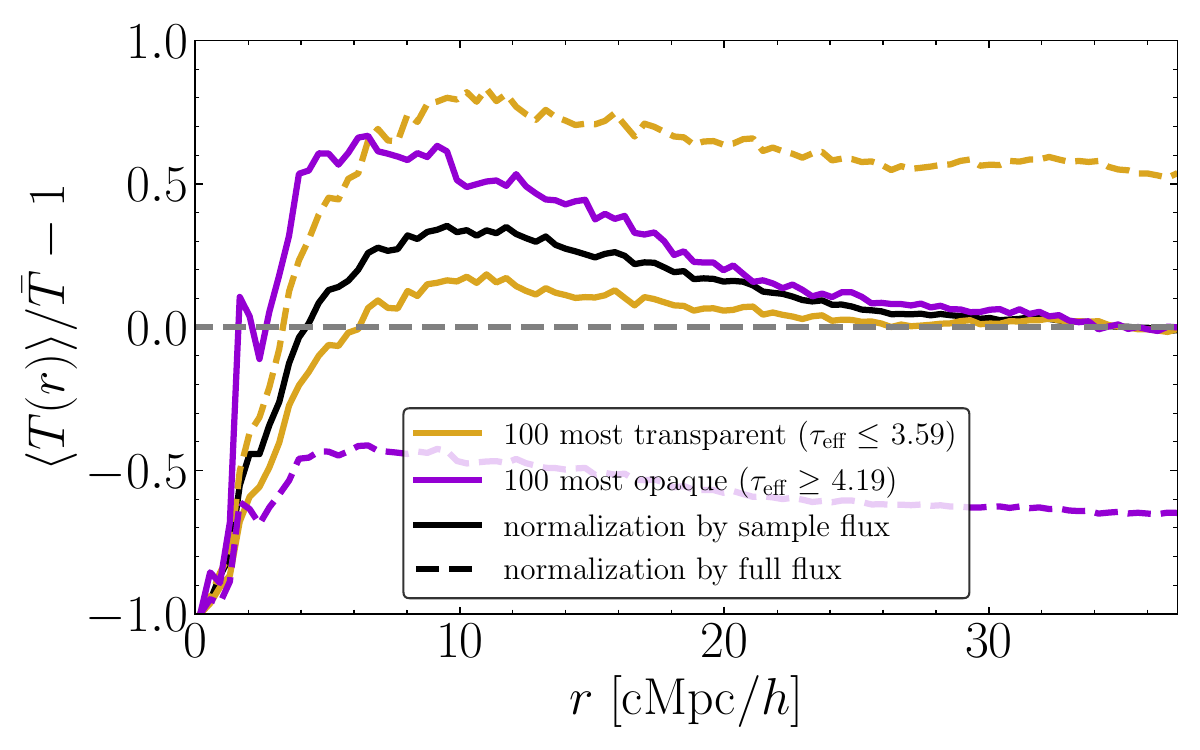}
    \caption{Impact of the normalization method chose for the \galacc. Solid lines are normalised by the average transmitted flux within the sample considered, while dashed lines are normalized by the average transmitted flux in the simulation box. The black line corresponds to a sample including all sightlines in the box, violet ones to the 100 most opaque sightlines and yellow ones to the 100 most transparent. The opaqueness of a sightline is estimated from its effective optical depth. 
    }
    \label{fig:corr_tau}
\end{figure}

We present in Fig.~\ref{fig:corr_tau} the \galacc computed for all sightlines in \thesanone (black solid line), using only the 100 most opaque ones (purple dashed line) and using only the 100 most transparent ones (yellow dashed line) and normalised by the average flux \textit{in the sample}. 
Comparing the dashed curves it appears clearly that the peak in transmissivity is larger in amplitude and located at smaller distances from galaxies when only the most opaque sightlines are considered. 
This might appear counter-intuitive given the interpretation of this as excess ionization (compared to the average) in the proximity of galaxies. 
However, this is a spurious effect of the normalization chosen. In practice, since each line of sight is not a fair sample of the distribution of densities and ionized fractions in the Universe, the normalization $\langle T(r) \rangle$ is larger (smaller) for the most transparent (opaque) sightlines, boosting the \galacc computed from the latter. This is shown explicitly in the Figure by the shaded regions 
that report the average transmitted flux as function of distance 
for the same two sightlines samples 
but now normalized to the average transmitted flux \textit{in the entire simulation} (and therefore sharing the same normalization). It now appears clear that transparent lines of sight show approximately 4 times more \Lya transmission than the opaque ones. At the same time, the peak is located to smaller distances from galaxies, likely as a consequence of the increased absorption of photons due to the more neutral environment, which constrain this ionizing radiation-driven proximity effect to smaller distances. We discuss this in more detail next.

\subsection{The relation between \galacc and mean free path}

The \galacc carries information on the local ionization field around the sources of reionization. In particular, the amplitude of the peak of transmitted flux is indicative of the strength of the ionizing emissivity of galaxies \citep[see][]{Kakiichi+2018} while its position is linked to the size of ionized regions around galaxies. The latter is, essentially, the mean free path of ionizing photons \textit{around} the galaxies used to compute the \galacc ($\lambda_\mathrm{mfp,gal}$). 

The connection between the \galacc peak position and the size of ionized regions, however, is not straightforward. In fact, the mean free path is sensitive to optical depths to ionizing photons $\tau \approx 1$, while the \galacc probes regions with $\tau \ll 1$. The latter stems from the fact that the \galacc is measured using the \Lya flux, which is completely absorbed whenever the (local) neutral fraction is $x_\mathrm{HI} \gtrsim 10^{-4}$. In other words, a photon escaping a galaxy will encounter much sooner a patch of the IGM with $x_\mathrm{HI} \gtrsim 10^{-4}$ than accumulating a total optical depth of $\tau \approx 1$, simply because it can easily cross the former without being absorbed, but not the latter (on average). Therefore, we expect the \galacc peak position to trace just the (size of the) innermost part of the ionized bubbles, where the neutral fraction is significantly below $10^{-4}$. This is further complicated by the evolution of the background UV density field. 

\begin{figure}
    \includegraphics[width=\columnwidth]{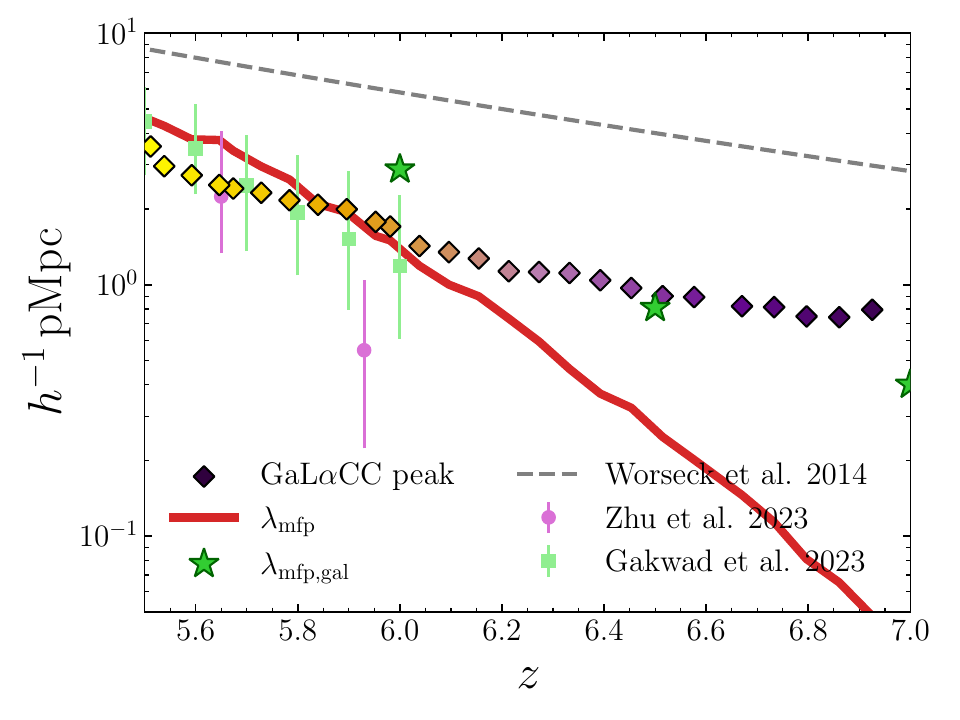}
    \caption{Comparison of the \galacc peak position (diamonds) and the mean free path of ionizing photons. For the latter, we report both the global value ($\lambda_\mathrm{mfp}$, solid line) and the local one ($\lambda_\mathrm{mfp, gal}$, green stars, computed around the most massive haloes in the simulations). 
    The dashed line shows the extrapolation of the fit \citet{Worseck+2014} computed from post-reionization mean free path measurements, while the green squares and purple circles report the measurements from \citet{Gaikwad+2023} and \citet{Zhu+23}, respectively. 
    }
    \label{fig:corr_mfp}
\end{figure}

In order to shine light on this issue, in Fig.~\ref{fig:corr_mfp} we compare the mean free path ($\lambda_\mathrm{mfp}$, solid red line, computed as described in \citealt{Thesan_igm}) and peak position evolution (diamonds, see Appendix~\ref{app:peak_evol_fit} for details on how this was computed, as well as for a fit to its redshift evolution). The peak positions has a much shallower evolution of the mean free path, which instead shows a rapid change between $z=7$ and $z=5.5$ as a result of the percolation of ionized bubbles at the end of cosmic reionization. This is a consequence of the fact that the \galacc probes regions around the (brightest) sources of ionizing photons, and therefore probes the first regions to be reionized, where the mean free path is expected to be the longest. We compare its evolution to the extrapolation of the fit from \citet{Worseck+2014}, computed from post-reionization mean free path measurements that are therefore affected only by the evolution of density fluctuations in the IGM. While the \galacc peak position is approximately $3.5$ smaller than the mean free path for the reason outlined above, their redshift evolution is strikingly similar. This suggests that the \galacc peak probes regions that are fully ionized already, in line with the physical interpretation of the \galacc (see the first paragraph of \eg Sec.~\ref{sec:results_obs}). We also show with green stars the mean free path around galaxies with $M_\mathrm{star} \geq 10^9 \, \Msun$ ($\lambda_\mathrm{mfp, gal}$) computed at $z=6, 6.5, 7$, showing. We find consistently that $\lambda_\mathrm{mfp, gal} > \lambda_\mathrm{mfp}$, as expected, and a redshift evolution intermediate between the extrapolation from \citet{Worseck+2014} and the one of $\lambda_\mathrm{mfp}$. The redshift evolution of $\lambda_\mathrm{mfp, gal}$ is however steeper than the one of the \galacc peak position, indicating that the evolution of ionized bubbles around the selected sources (traced by the local mean free path) is faster than the evolution of the the highly-ionized region probed by the \galacc. For comparison, we also report the recent measurements of the mean free path from \citet[][green squares]{Gaikwad+2023} and \citet[][purple circles]{Zhu+23}.

\subsection{Local gas properties}
\label{subsec:local_gas}

\begin{figure}
    \includegraphics[width=\columnwidth]{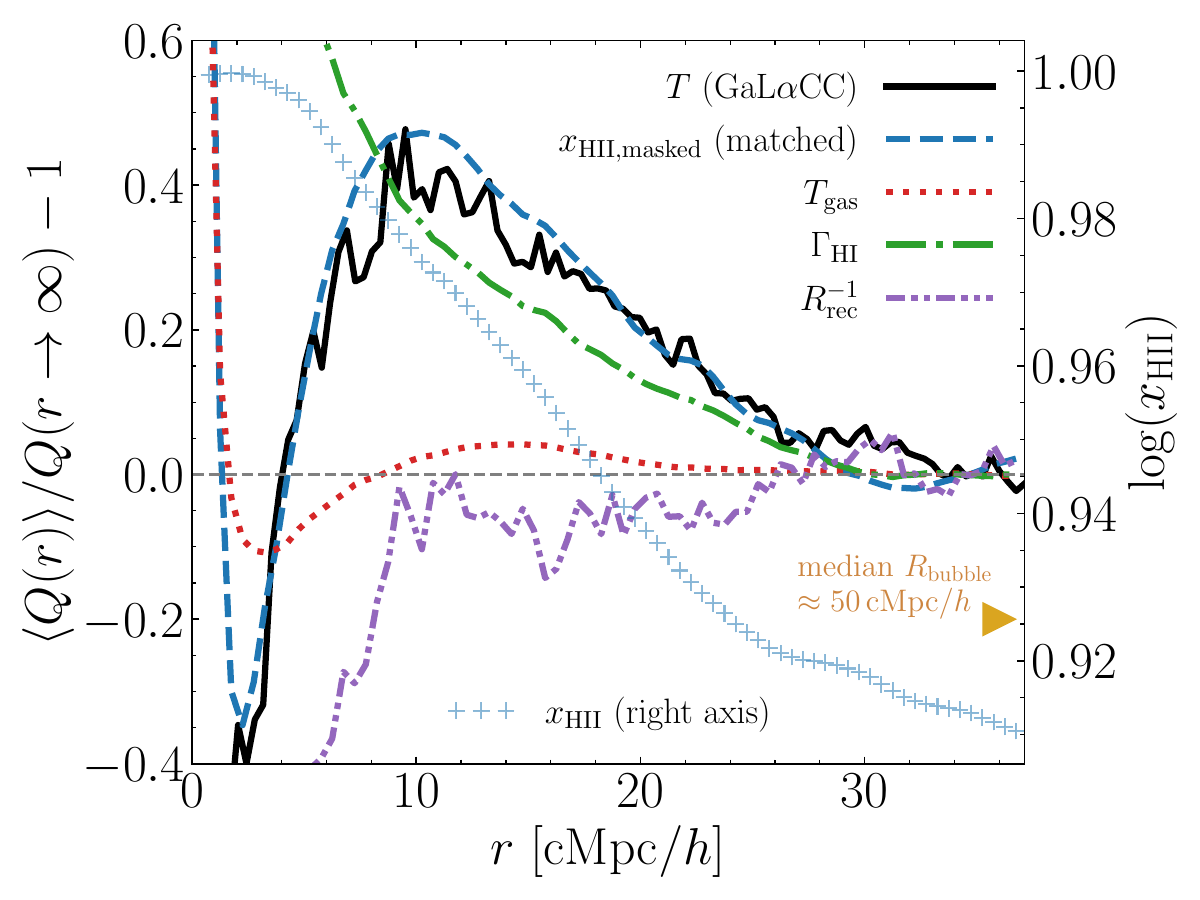}
    \caption{Mean radial profile of different quantities $Q$ around all galaxies in the \textsc{Thesan-1} simulation with stellar mass $M_\mathrm{star} \geq 5 \times 10^8\,M_\odot$. We show two \HII profiles, namely the full one (blue crosses, to be read against the right-hand-side vertical axis) and one computed only considering gas cells with local neutral fraction $x_\mathrm{HI} \leq 10^{-4}$ (dashed blue line), to account for the fact that above such threshold the \Lya flux is completely suppressed. The latter is also manually rescaled in the vertical direction to approximately match the \galacc curve (black solid line), highlighting that the latter traces very well the profile of \textit{extremely ionised gas}, but not of all gas. The median bubble size in the simulation is $\approx 50$ cMpc/$h$ \citep{Thesan_bubbles}, much larger than the scales probed by the \galacc.
    }
    \label{fig:profiles}
\end{figure}

Finally, we investigate the local gas properties around the sources and their role in setting the \galacc. To this end, we compute the volume-averaged radial profile of different simulated quantities $Q$ around all galaxy with stellar mass $M_\mathrm{star} \geq 5 \times 10^8\,\Msun$). We chose this stellar mass threshold because it provides a visually-cleaner picture, but we have checked that results are robust against it. 
We show the resulting profiles in Fig.~\ref{fig:profiles}, normalised by their asymptotic ($r \rightarrow \infty$) value minus one, for the sake of visual clarity. The quantities shown are: \Lya transmissivity (black solid line),  HI photoionisation rate ($\Gamma_\mathrm{HI}$, dot-dashed green line), gas temperature ($T_\mathrm{gas}$, dotted red line), HI fraction (blue crosses, to be read against the right-hand-side vertical axis), the HI fraction computed only for gas cells with ($x_\mathrm{HI} \geq 10^{-4}$ to mimic the \Lya visibility, since gas less ionised than this threshold will completely suppress the \Lya flux), inverse of the recombination rate ($R_\mathrm{rec}^{-1}$, double-dot-dashed purple line). 
Note that when the quantity investigated is the \Lya transmitted flux (\ie $Q=T$), we obtain the \galacc.

The behaviour of the \galacc and its relation to the local gas properties is cleanly split at its peak location. On larger scales, the \Lya transmission excess follows very closely the $\Gamma_\mathrm{HI}$ excess profile, increasing as the distance form the sources decreases. In this region, the HI recombination rate is approximately constant. Therefore it does not affect the \textit{modulation} of the \Lya transmission with spatial scales. The situation dramatically changes at the \galacc peak scale. At this and smaller scales, the recombination rate sharply increases (thus its inverse, shown in the plot, decreases). This is a because at these scales the environment of the overdensities sourcing the \galacc (see Sec.~\ref{subsec:delta}) becomes important and the enhanced gas density boosts recombination into HI, therefore suppressing the \HII fraction and thus the \Lya flux.

This competition between the HI photoionization rate and recombination rate effectively modulates the amount of \textit{extremely ionised} hydrogen, which determines the \Lya flux. We stress that, given the large oscillator strength of \Lya, only gas with local neutral fraction $f_\mathrm{HI} \lesssim 10^{-4}$, since the gas is otherwise fully opaque to \Lya regardless of the gas conditions. We explicitly show this in Fig.~\ref{fig:profiles} by plotting two version of the $x_\mathrm{HII}$ profile around galaxies. One is computed using all gas (blue crosses, to be read against the right-hand-side vertical axis) and shows that the \HII fraction is monotonically decreasing with distance, but remains above 90\% until $r \approx 50\,\mathrm{cMpc}/h$, showing no correlation with the \galacc. However, when all the \Lya opaque gas is masked out (by only considering gas with local neutral fraction $f_\mathrm{HI} \leq 10^{-4}$, shown using a blue dashed line and to be read against the left-hand-side verical axis) the profile is strikingly similar to the \galacc. Note that we have rescaled this curve to visually match the \galacc in amplitude. This is necessary because the variations in the \HII fraction are tiny (approximately $10^{-5}$) compared to those in the \Lya flux due to the very large oscillator strength of this line and would not be visible in the plot otherwise. 
This clarifies further that the \galacc can not be directly compared to ionised bubbles sizes, since the latter are defined typically using $\mathcal{O}(1)$ variations in ionised fraction. For instance, in the \textsc{thesan-1} simulation at the redshift investigated the median bubble size is $R_\mathrm{bubble, median} \approx 50\,\mathrm{Mpc}/h$ \citet{Thesan_bubbles}, much larger than the \galacc peak and, in fact, at such scale the \Lya transmissivity is already at its average value. It should be noted, however, that box size limitation might quantitatively affect this, since gas quantities are forced to converge to their average value at scales $r \lesssim L_\mathrm{box} \sqrt{3}/2$, while the bubble size is computed along individual lines of sight, that do not suffer from this limitation. This has also been hinted at recently in \citep{Conaboy+2025}.

We also investigate the role of gas temperature in shaping the \galacc. The radial profile of this quantity shows two salient features. At scales larger than few cMpc/$h$, it shows a modulation somewhat similar to the \galacc, although the peak is much smaller in amplitude and at somewhat larger scales. At smaller scales, it exhibits a sharp increase that reaches orders of magnitude above the asymptotic value. This is a consequence of feedback processes within the galaxies that inject large amounts of heat in their surroundings. For the same reason, the \HII fraction is also boosted by similar amounts. Overall, these discrepancies points to a negligible impact of gas temperature in setting the \galacc at this redshift. We verify this by producing synthetic spectra with gas temperature fixed to $3 \times 10^4$ K and find virtually no difference in the \galacc. Note, however, that $T_\mathrm{gas}$ might play a larger role at lower redshift, as reported recently by \citep{Conaboy+2025}.

\section{Discussion and future perspective}
\label{sec:future}
Now that we have thoroughly characterized the \galacc sensitivity to a number of observational effects and further clarified its physical origin, we move on to discuss the implications of such knowledge and investigate future developments it inspires. 

\subsection{Which surveys can be used to measure the galaxy--Lyman-$\alpha$ cross-correlation?}
Leveraging our results presented in Sec.~\ref{sec:results_obs}, we can outline which current and planned surveys might be able to faithfully measure the \galacc either using available data or with additional observational efforts. In Table~\ref{tab:surveys} we provide a list of ongoing or planned surveys. For each of them we list the expected number of spectra, their spectral resolution, the S/N and the area covered around each line of sight. 
It reveals that a number of ongoing surveys can potentially be used to greatly extend the measure of the \galacc. For instance, the WEAVE-QSO survey appears to be a promising candidate to extend observations of this quantity if complemented with observations of the fields around the QSO (at the moment there are no WEAVE-IFU observations planned for these fields; Welsh, private comm.), or by identifying galaxies through their absorption features in the spectra themselves. 
However, its sensitivity may not be sufficient to detect galaxies at the redshift discussed in this paper. 
Similarly, the `High-Redshift Sample' obtained as part of the 4MOST-QSO survey \citep{4MOST-QSO} is poised to obtain a large number of spectra of $z\geq6$ QSOs at intermediate spectral resolution but without information on the surrounding fields. Should this information be available through complimentary surveys, the potential of the \galacc could be fully exploited. Finally, the combination of DESI and of its legacy survey can deliver an accurate measurement of this cross-correlation, although the impact of the Legacy Surveys completeness for galaxies at $z\sim6$ has to be assessed before any definitive claim can be made. 

Overall, our work shines an optimistic light on the possibility to use ongoing or forthcoming surveys for the measurement of the \galacc. Many of such QSOs are too faint for traditional \Lya studies (which require high S/N to resolve individual features, making their observations prohibitively expensive for faint objects), but the resilience of the \galacc to noise (see Sec.~\ref{subsec:noise_specres}) entails that they could potentially be used for this measurement. In addition, a number of new-generation instruments are currently being deployed or developed. For instance MOONS on VLT \citep{MOONS}, SUBARU-PFS \citep{PFS} and MOSAIC on the future ELT \citep{MOSAIC} have promising capabilities for observing the \galacc, thanks to their high multiplexing and resolving power. In the more distant future, an instrument like the Wide-field Spectroscopic Telescope \citep{WST} could make the \galacc a routine observation.

\begin{table}
    \caption{Relevant properties of some ongoing/planned surveys, colored in green, orange or red to indicate whether they are well suited, sufficient or insufficient (respectively) for computing the \galacc.
    }
    \label{tab:surveys}
    \centering
    \renewcommand{\arraystretch}{1.333333333333333}
    \begin{tabular}{l|c|c|c|c}
         \textbf{survey name} & \textbf{number of} & \textbf{spectral} & \textbf{pixel} & \textbf{area per LOS} \\
                              & \textbf{spectra} & \textbf{resolution} & \textbf{S/N} & \textbf{(arcmin$^2$)} \\
         \hline
         EIGER$^{*}$           & \textcolor{BurntOrange}{6}                    & \textcolor{ForestGreen}{$\sim 9000$}    & \textcolor{ForestGreen}{$\lesssim 200$} & \textcolor{ForestGreen}{$6.5 \times 3.4$} \\
         ASPIRE$^{\triangle}$  & \textcolor{ForestGreen}{25}                   & \textcolor{ForestGreen}{$\sim 9000$}    & \textcolor{ForestGreen}{$\lesssim 38$}  & \textcolor{ForestGreen}{$\approx 11.2$} \\ 
         
         WEAVE-QSO$^{\circ}$   & \textcolor{BurntOrange}{$\lesssim 10$}        & \textcolor{BurntOrange}{$\sim 5000$}    & \textcolor{BurntOrange}{$\lesssim 5$} & \textcolor{BrickRed}{0}  \\
         4MOST-QSO$^{\dagger}$ & \textcolor{ForestGreen}{$\sim 4 \times 10^4$} & \textcolor{BurntOrange}{$4000$--$7000$} & N/A & \textcolor{BrickRed}{0}  \\
         DESI$^{\ddagger}$     & \textcolor{ForestGreen}{$\gtrsim 150$}        & \textcolor{BurntOrange}{$3000$--$5000$} & N/A & \textcolor{ForestGreen}{$\gg 100$}  \\
    \end{tabular}
    \renewcommand{\arraystretch}{0.75}
    \begin{flushleft}
                 $^{*}$\citet{EIGERI, EIGERII}
        \newline $^{\triangle}$\citet{aspire}
        \newline $^{\circ}$\url{https://ingconfluence.ing.iac.es/confluence//display/WEAV}; as of now, there is no plan to observe the WEAVE-QSO fields with WEAVE-IFU (Welsh, Pieri, private comm.)
        \newline $^{\dagger}$High-Redshift sample, \citet{4MOST-QSO}, \url{https://www.eso.org/sci/facilities/develop/instruments/4MOST.html}
        \newline $^{\ddagger}$ including the Legacy Surveys, \citet{DESI-highzQSO}
    \end{flushleft}
\end{table}

\subsection{Can we use galaxies as background sources?}
Our finding that even noisy spectra, when used in sufficiently large number and/or with sufficiently high spectral resolution, can faithfully recover the cross-correlation signal (Sec.~\ref{subsec:noise_specres}) opens up the possibility to use galaxy spectra instead of QSO ones as background sources. Thanks to their larger number density, this would open up a swath of new possibilities, including the study of spatial variations in the \galacc, extending its study to higher redshift and even directly mapping the impact of ionizing photons from individual bright objects. 
This would represent an extension of the so-called IGM tomography, \ie the reconstruction of the large-scale post-reionization density field in the IGM using the \Lya absorption in the spectra of background galaxies (typically Lyman break galaxies) as an indicator of the density \citep[\eg][]{clamato,clamato2,Newman+2020,Ravoux+2020,Kakiichi+2023}. Forthcoming instruments like PFS\footnote{\url{https://pfs.ipmu.jp/index.html}} and MOSAIC\footnote{\url{https://elt.eso.org/instrument/MOSAIC/}} are poised to significantly extend our ability to perform such studies. 

Measuring the \galacc using background galaxies presents a number of challenges, among which stands out the fact that galaxy spectra are much fainter and more intrinsically variable than QSO ones, both conspiring to make their continuum harder to reconstruct, and therefore the measurement of their transmissivity less precise. Importantly, this has the potential to break one key assumption of our results discussed above, namely that the noise is uncorrelated at the pixel level. 
In order to progress forward without losing generality, we take a simplified approach. Specifically, we assume errors in the spectra reconstruction result in two types of somewhat idealized modification to the intrinsic normalised flux. While this reduces the direct applicability of our results to real observations, overcoming them would require us to make strong assumption on the type of instrument used to collect the galaxy spectra, as well as on the continuum estimation procedure. For this reason, in this work we choose to remain agnostic to such details and provide a proof of concept analysis of such residual correlated noise, which we hope can guide the design of future observations. 

\begin{figure*}
    \centering
    \includegraphics[width=0.9\textwidth]{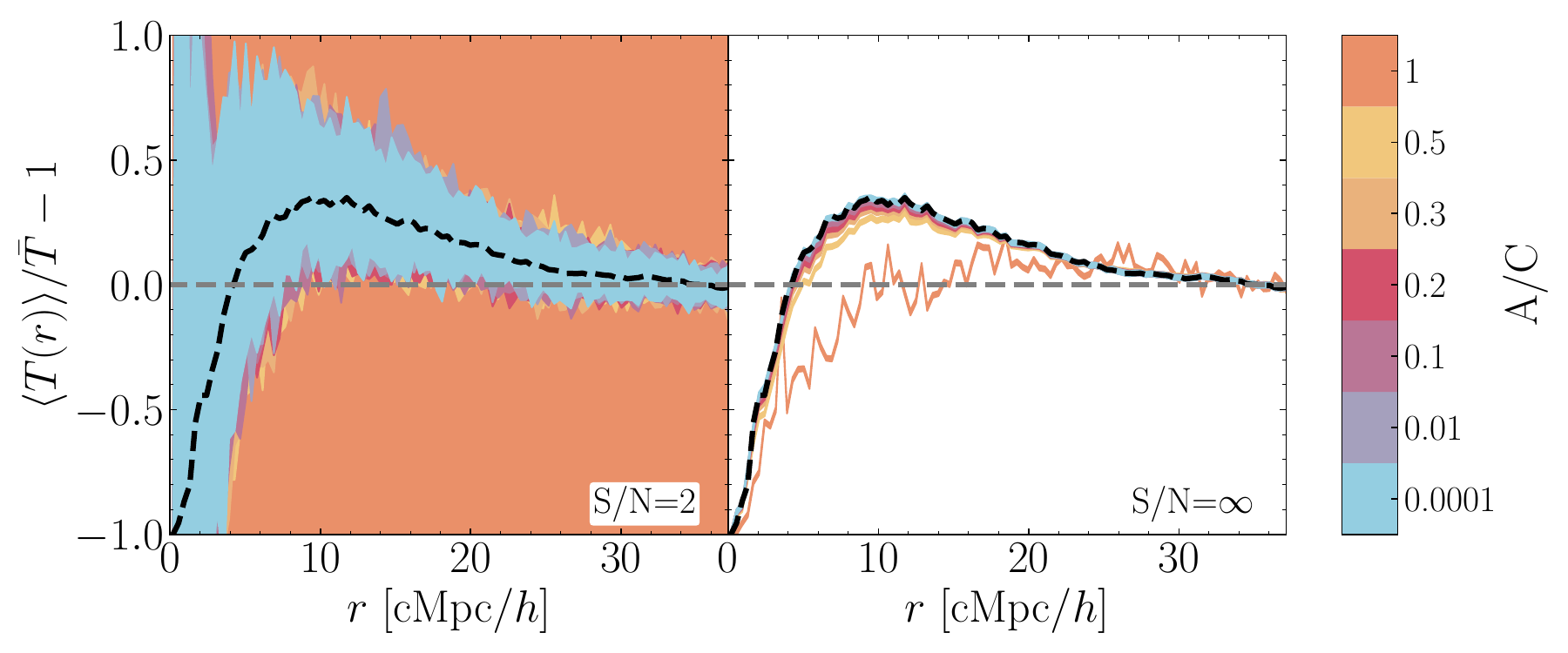}
    \caption{Impact of correlated noise due to residuals in the source continuum subtraction for spectra with $R=4000$ and $S/N=2$ (left panel) and $S/N=\infty$ (right panel). The (very simple) correlated noise model is described in the text. The colored bands show the envelope of 50 different noise realizations for different values of the residual correlated noise amplitude $A/C$ (see text for a precise definition). It can be seen that only when the residual noise reaches the continuum level (\ie $A/C \sim 1$), the correlation signal is lost. In the left panel we show with a dashed black line the synthetic noise-free $R=300\,000$ cross-correlation.
    }
    \label{fig:corr_noise_modulation}
\end{figure*}

\begin{figure*}
    \centering
    \includegraphics[width=0.9\textwidth]{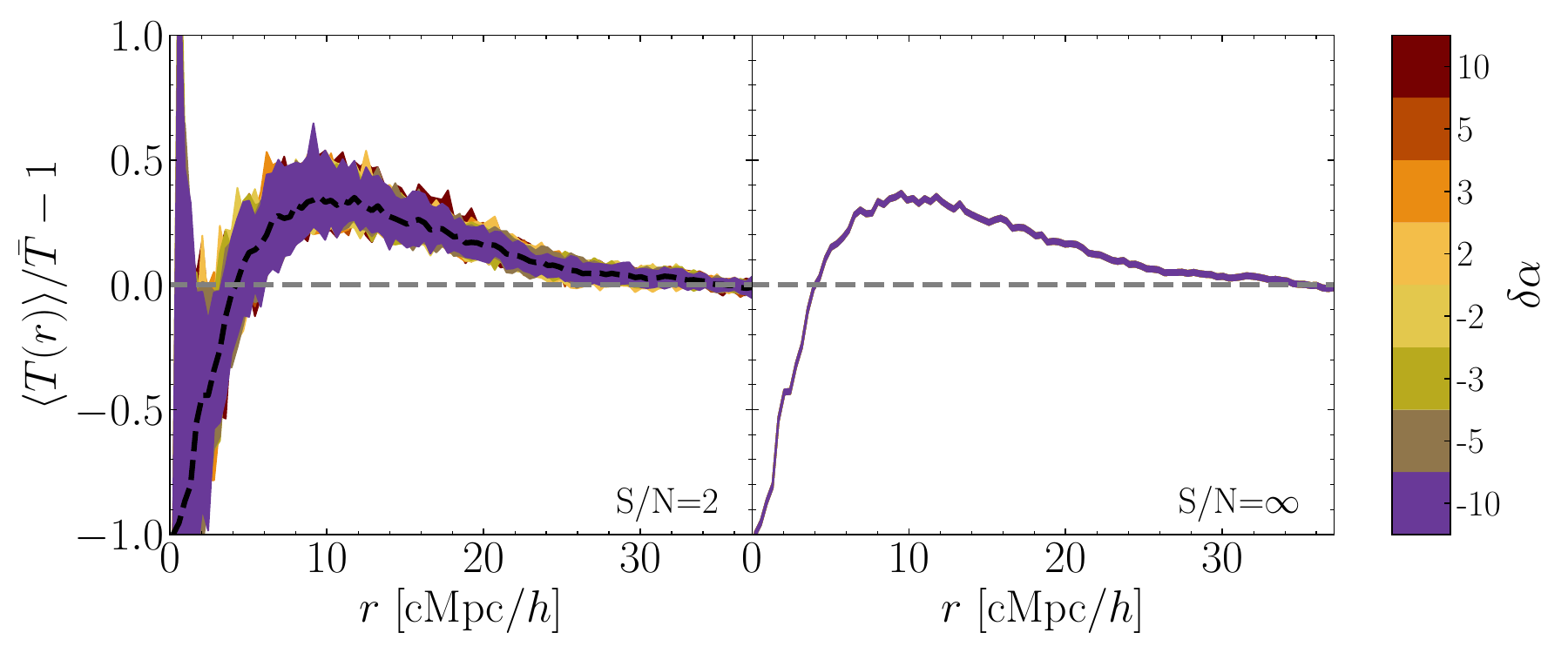}
    \caption{Same as Fig.~\ref{fig:corr_noise_modulation} but for errors ($\delta \alpha$) in the estimated continuum slope. }
    \label{fig:corr_slope_error}
\end{figure*}

In the first type of reconstruction errors, we assume that the reconstructed normalised flux $\Tilde{f}$ is the superposition of the true normalized flux $f$ and a sinusoidal signal. This could result, for example, from unmodeled variability in the galaxy continuum when it is assumed to be a power-law. Starting from synthetic noise-free spectra with resolution $R=4000$ (taken as representative of the spectral resolution of the MOONS, PFS and MOSAIC instruments), we modify the transmitted flux $f \equiv F/C$ (where $F$ is the observed flux in a pixel of the spectrum and $C$ is the estimated continuum in the same pixel) as:
\begin{equation}
    \Tilde{f} \equiv \frac{F}{C + A \sin(n \lambda)} = \frac{f}{1+\frac{A}{C}\sin(n \lambda)} \, ,
\end{equation}
where $A$ is the amplitude of the residual noise, and $n$ controls the frequency of the modulation. We have checked that the results do not depend on  the latter, that we therefore fix to $n=1$ (remember that this analysis is qualitative from its inception, and therefore we are not worried about small quantitative differences due to $n$). On top of this residual correlated noise, we add a Gaussian (uncorrelated) noise to achieve a pixel-level average $S/N=2$. 

The second type of reconstruction error we study stems from errors ($\delta \alpha$) in the estimation of the true continuum slope $\alpha$, resulting in $\Tilde{f}$ deviating more and more from $f$ as a function of wavelength $\lambda$. Specifically, we assume that the true continuum $C = C_0 \lambda^\alpha$ is correctly estimated at the \Lya wavelength ($\lambda_{\mathrm{Ly}\alpha}$). Therefore, we model the reconstructed normalised flux as: 
\begin{equation}
    \Tilde{f} \equiv \frac{F}{C_0 (\lambda/\lambda_{\mathrm{Ly}\alpha})^{\alpha + \delta \alpha}} = f \left( \frac{\lambda}{\lambda_{\mathrm{Ly}\alpha}} \right)^{-\delta \alpha} \, .
\end{equation}

We show the inferred \galacc for the two types of reconstruction error described in Fig.~\ref{fig:corr_noise_modulation} and Fig.~\ref{fig:corr_slope_error}. The former shows the impact of correlated noise in the reconstructed continuum by varying the value of $A/C$, while the latter reports the impact of errors ($\delta \alpha$) in the estimated continuum slope. In both Figures, the left and right panels report the results assuming a pixel-level S/N$=2$ and S/N$=\infty$ (\ie noise free spectrum), respectively. The figures clearly show that the investigated reconstruction errors do not significantly affect the inferred \galacc, as long as they are not extreme ($A/C \sim 1$ for the correlated noise and $\delta \alpha > 10$ for the slope errors). 

This finding is very promising, as it seems to indicate that even large errors in the continuum reconstruction (as expected for galaxies) do not prevent the recovery of the intrinsic \galacc. However, we expect this conclusion to depend, at least quantitatively, on the specific details of the spectra (\eg spectral resolution, total length) and on the galaxy sample employed. We leave this for a future work, as it requires to perform survey-specific forecasts that would break the generality we strive for in this paper.

\subsection{Extension to higher Lyman series lines}

\begin{figure}
    \includegraphics[width=\columnwidth]{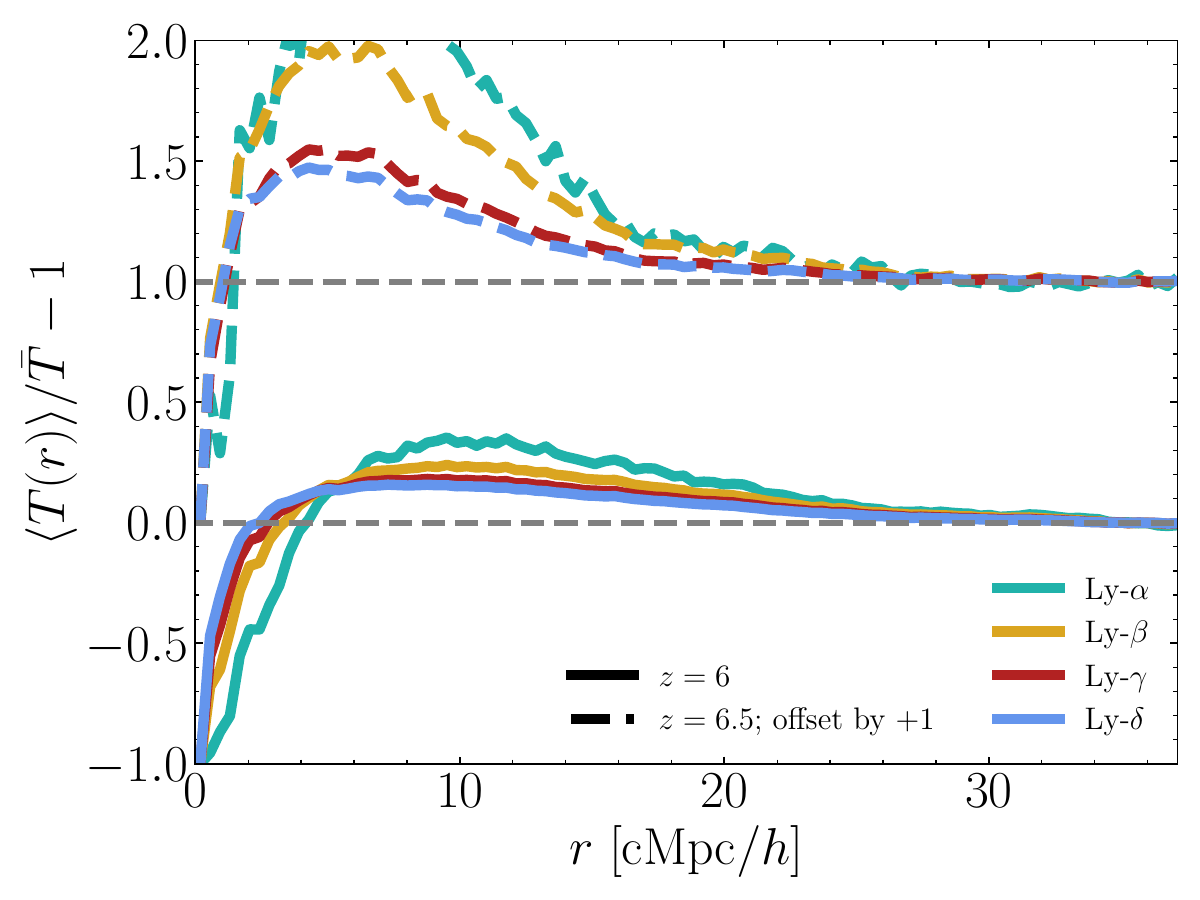}
    \caption{Galaxy--Lyman series cross-correlation at $z=6$ (solid lines) and $z=6.5$ (dashed lines, offset by 1 for visual clarity). Different colours correspond to different Lyman series lines, as reported in the caption. The grey dashed line shows the zero level, corresponding to the average transmitted flux in the Universe.}
    \label{fig:corr_Lyseries}
\end{figure}

Another possible development of the \galacc is the replacement of the \Lya line with higher-order Lyman-series lines. Thanks to their lower oscillator strengths, they are less easily saturated, enabling their detection in less ionized regions, and thus higher redshift. However, they are more and more easily contaminated by lower-redshift \Lya transmission, limiting the spectral region where they can easily and reliably be identified (for example, this region for Lyman-$\beta$ is approximately a third of the one for \Lya). Nevertheless, the use of Lyman-$\beta$ flux is increasingly common in IGM studies at the tail end of reionization \citep[\eg][]{Eilers+2019, Jin+2023, Ding+2024}. This is promising to enable higher-redshift studies of the \galacc. Moreover, thanks to their larger saturation densities, higher-order Lyman series lines enable us to probe closer to the overdensities where galaxies reside. 

As a starting point for a future thorough study of this, we present in Fig.~\ref{fig:corr_Lyseries} the Galaxy--Lyman series cross-correlation at $z=6$ (solid lines)  and $z=6.5$ (dashed lines, offset by 1 for visual clarity) for the \Lya (green line), \Lyb (yellow line), \Lyc (red line) and \Lyd (blue line). The curves share the overall shape (discussed in Sec.~\ref{sec:results_obs}), but with some differences. Higher-order lines show a smaller and closer-to-galaxies excess of transmitted flux, and the latter is suppressed at smaller scales. This is true at both redshifts shown in the Figure (and we have ensured that it is true for all redshift in the range $5.5 \leq z \leq 7$), although it is more evident at later times. The shift in peak position and in the region of suppressed flux stem simply from the fact that the \Lyb, \Lyc and \Lyd transmission require progressively larger neutral fractions to be entirely absorbed, which are achieved in progressively smaller regions around galaxies. The peak amplitude, instead, is reduced because the average transmitted flux for higher-order transitions increases more in the `background' (\ie far from the galaxies) than close to the sources. This is due to the increased difficulty of boosting the transmitted flux in already highly-ionized regions, as it is the case close to the sources of ionizing photons) with respect to not-so-ionized ones. Overall, our results indicate that higher-order Lyman transitions can extend to higher redshift our ability to study the galaxy--IGM connection through the \galacc, as well as to probe regions closer to the galaxy. In fact, it is conceivable to combine the cross-correlation of different Lyman-series line to statistically probe the density field around high-redshift galaxies. We leave a detail study of this to a future work.

\subsection{Measuring the \galacc for Helium reionization}
Finally, we address one last question concerning the \galacc, namely the possibility to extend the use of this probe to the Helium  epoch of reionization (HeEoR), replacing the \HI \Lya line with the \HeII one and galaxies with QSOs. There are a number of factors, both contributing and hindering this endeavor, to take into account. For this reason, a full examination of this problem requires a dedicated work, that we plan to deliver in the future. However, we briefly discuss here the main considerations and insights from our results.

The HeEoR occurs at much lower redshift than the hydrogen reionization discussed so far \citep[approximately at $z\sim 3.5$, see \eg][]{worseck2016,davies2017,worseck2019,makan2021,makan2022}, where the number density of QSO is higher and, therefore, there are potentially more sightlines that can be used to build the cross-correlation. For example, WEAVE \footnote{The WHT Enhanced Area Velocity Explorer, \url{https://ingconfluence.ing.iac.es/confluence//display/WEAV}} will soon more than double the number of sightlines observed at $z>2$ (covering several thousand square degrees). Although the wavelength range of WEAVE does not cover the redshifted \HeII \Lya wavelength of \mbox{$304$ \AA $(1+z) \sim 1200$ \AA} (for $z\sim 3$), it can be used to identify sightlines clean from contamination of higher-$z$ \HI Lyman continnum for follow up. At the same time, eBOSS\footnote{\url{https://www.sdss4.org/surveys/eboss/}}, 4MOST\footnote{\url{https://www.eso.org/sci/facilities/develop/instruments/4MOST.html}} and EUCLID\footnote{\url{https://sci.esa.int/web/euclid/}} are also expected to significantly increase the number of known QSOs (although, again, without the necessary wavelength coverage). However, observations of the \HeII \Lya forest are significantly more challenging than those of the \HI counterpart, due to the aforementioned contamination and to the necessity of spaced-based observations (since the atmosphere is opaque at the FUV wavelengths where the helium forest is found). Additionally, the rarity of the sources driving the HeEoR implies that a much 
larger area around each sightline needs to be covered in order to identify sources contributing to the \galacc \citep[since they produce much larger ionized bubbles, \eg][]{Compostella+2013}. 
The situation is worsened by the fact that QSO emission is thought to be strongly anisotropic, which introduces geometrical (unobservable) weights to the contribution of each source around the line of sight. However, QSOs are also much brighter than galaxies, therefore imprinting a much stronger signature on the surrounding \Lya forest which might make up for the larger sensitivity to source-to-source variations. Finally, QSOs reside in the largest overdensities \citep[\eg][]{Efstathiou&Rees1988,Volonteri&Rees2006,Costa+2014,Costa2024}, therefore boosting the \HeIII recombination rate and suppressing the \HeII \Lya transmissivity. 
Overall, this appears a complex problem deserving a dedicated study in the future. 

While the source-\Lya cross-correlation has never been studied in the HeEoR, there have been somewhat similar attempts. For instance, \citet{Schmidt+2017} presented a survey of QSOs at $z\sim3$ designed specifically to detect the \HeII transverse proximity effect, \ie the enhancement in the nearby \Lya transmissivity due to the quasar ionizing photon production, that resulted in 66 `science-grade' spectra probing 4 different foreground QSOs. Their goal was to constrain the lifetime and obscuration of individual QSOs, and therefore analyzed individual features in the spectra \citep[][finding evidence of such proximity effect]{Schmidt+2018}. However, a similar, expanded sample would be a promising tool to measure the source-\Lya cross-correlation and put new constraints on the HeEoR, pushing forward one of the least studied periods in the history of structure formation. This acquired even more importance in recent times, since it was pointed out by \citet{Basu+2024} that our current model of QSO luminosity function evolution yields results that are somewhat in tension with the (still sparse) constraints on the HeEoR. 

\subsection{Implications for the relation between sightline effective optical depth and galaxy density}
The \galacc is closely related to the relationship between the total \Lya effective optical depth of a sightline and the number density of galaxies around it, that has been recently measured in $7$ quasar sightlines \citep{Becker+2018,Kashino+2020,Christenson+2021, Christenson+2023, Ishimoto+2022}, including some of the most transparent and most opaque LOS known at $z \sim 6$--$7$. We investigate the implications of the \galacc and the predictions of \thesan regarding this quantity in a companion paper \citep{Thesan_igm3}.

\section{Summary and Conclusions}
\label{sec:conclusions}

The galaxy--\Lya cross-correlation (\galacc) has been proposed as a novel way to constrain the properties and timing of cosmic reionization. In this paper we have (\textit{i}) thoroughly investigated the extent to which a different observational limitations prevent us from extracting the intrinsic \galacc signal, as well as (\textit{ii}) further clarified the physical origin of the signal and its dependence on galaxy properties and their environment. In order to do so, we have employed the \thesan suite of simulations (specifically, the \thesanone flagship run) to produce synthetic \Lya forest lines of sight. Our main results are the following:

\begin{itemize}
	\item The intrinsic sightline-to-sightline variation in the \galacc implies that a large number of sightlines is needed to overcome such structure formation noise and recover the underlying signal. We found that, ideally, the total length of the spectra used to compute the \galacc should exceed $L_\mathrm{tot} \gtrsim 2500 \, \mathrm{cMpc}$. A total length equal to half this value appears to already provide a reasonable fidelity in most sightlines, although the possibility of nevertheless obtaining a result far away from the intrinsic signal can not be neglected beyond reasonable doubt in this case. The exact number of spectra depends on the length of the redshift bin used.
	\item We predict the observations by \citet{EIGERI} to be dominated by stochasticity, and therefore we caution against interpreting them as representative (qualitatively or quantitatively) of the global \galacc signal. On the contrary, we predict that the latest EIGER results \citep{EIGERVII}, including all 6 sightlines observed, are starting to reliably probe the true \galacc signal and, therefore, typical IGM conditions. Similarly, the recent results by the ASPIRE program \citep{ASPIRE_galacc} span spectral length large enough to beat cosmic variance, but their shallower observations naturally yield a noisier signal.
	\item The galaxy selection method appears to be largely irrelevant for the \galacc. We do not find differences when the cross-correlation is computed using the galaxies with the largest stellar mass, SFR, gas metallicity (as a proxy for their \CIV absorption) and \OIII flux. This stems from the fact that \thesan predicts galaxies at $z\gtrsim 6$ to lie on the galaxy main sequence and the mass-metallicity relation, and the \OIII flux to be correlated with the galaxy SFR. Hence, galaxies selected to be extreme in one of these quantities are typically extreme also in the others. 
	\item The spectral resolution and signal-to-noise ratio of the spectra have a limited impact on the \galacc, thanks to its statistical nature. This, however, quantitatively depends on the number of spectra used. Extremely poor spectral resolution or S/N degrade the predictions beyond reliability regardless of the total number of spectra.
	\item The excess transmitted flux at intermediate scales is driven by galaxy overdensities rather than individual bright objects. Therefore, the \galacc does not trace ionized bubbles produced by individual galaxies, but rather large-scales collective bubbles. This clarifies why \citet{Thesan_igm} found that, simultaneously, (\textit{i}) even when ionizing photons are allowed to escape only from small ($M_\mathrm{halo} \leq 10^{10} \, \Msun$) galaxies, the \galacc signal is not affected, and (\textit{ii}) the flux excess is larger around massive ($M_\mathrm{star} \geq 10^9 \, \Msun$, $M_\mathrm{halo} \gtrsim 10^{11} \, \Msun$) galaxies (\ie when only such galaxies are used to compute the \galacc). The emerging picture is the following: small galaxies provide the majority of the ionizing photons required to induce an excess \Lya transmission, but individual sources are not able to produce an observable signature because of the large `structure formation noise'. Therefore, only large groups of such galaxies can produce an observable signature. These groups typically reside in overdense regions, as do highly biased sources like massive galaxies. Hence, using these massive objects to compute the \galacc effectively selects conglomerates of small galaxies, producing a stronger signal. At the same time, if larger galaxies are not allowed to emit ionizing photons, this does not significantly affect the \galacc, since the majority of ionising photons is produced by the numerous, nearby smaller galaxies. 
    \item The most opaque (transparent) sightlines to \Lya photons show larger and closer to the galaxy (smaller and farther from the galaxy) peaks in the \galacc. This is due to: (\textit{i}) the fact that the transmitted flux is normalized by its average in the sightline, which is larger in transparent sightlines; and (\textit{ii}) the overall lower ionising photons density in opaque sightlines renders more difficult to reach the high ionization levels required to allow \Lya transmission, and therefore moves the peak transmissivity at smaller distances from the galaxy. 
	\item We produced sightlines including lightcone effects and compared their predictions to the one obtained at fixed redshift. We find that they are largely equivalent as long as the redshift bin employed is below $\Delta z \lesssim 0.4$, although we expect the exact number to depend on the speed of reionization at the redshift of the observations. For larger redshift windows, we empirically determine the redshift of the best-matching fixed-redshift \galacc and show that such effective redshift is very close to the median redshift of the galaxies employed to compute the cross-correlation.
    \item The redshift evolution of the \galacc peak position is compatible with the evolving density field in a fully-ionized IGM, as expected around the brightest sources of reionization. Quantitatively, the position is approximately $3.5$ times smaller than the mean free path evolution extrapolated from \citet[][which is computed from measurements in the post-reionization IGM]{Worseck+2014}.
	\item We provide an initial investigation of the possibility to replace quasars with galaxies as background sources. This is particularly relevant with the rise of integral-field units and multi-object spectrographs, enabling us to obtain a large number of medium-resolution spectra simultaneously. Our results, despite the simplicity of this initial study, are promising. The recovered \galacc is not affected by error in the continuum reconstruction, provided the errors are below 100\% of the continuum value. It remains to be seen how resilient this is to decreasing spectral resolution and S/N.
	\item We review the specification of a number of ongoing surveys (namely, EIGER, ASPIRE, WEAVE-QSO, 4MOST-QSO and DESI) and determine which one possess the specification required to compute the \galacc. We find that WEAVE-QSO and 4MOST-QSO are promising candidates for this task, if they can be extended with additional data on the QSO fields. DESI, on the other hand, could already provide all the necessary data, but the impact of its Legacy Surveys completeness for galaxies at the end of reionization needs to be thoroughly assessed.
\end{itemize}

Overall, our analysis sheds new light on the galaxy--\Lya cross correlation. It clarifies further the physical origin and the thoroughly investigate the observational requirements for a faithful determination of this signal, reviewing the ability of current and forthcoming surveys to deliver a faithful measurement. We also provide an extensive discussion of future possible extensions of these type of studies. As such, this paper represents a step further to a deeper understanding of the galaxy--IGM interplay during the first billion years of the Universe, a step direly needed to shed light on the many questions being on a daily basis by observations of the reionizing Universe.

\section*{Acknowledgements}
EG wishes to express his gratitude towards Koki Kakiichi and Christopher Cain for discussions that inspired part of this work, as well as towards Louise Welsh, Daichi Kashino, Valentina D'Odorico and Matthew Pieri for useful discussions. We are thankful to the community developing and maintaining software packages extensively used in our work, namely:  \texttt{matplotlib} \citep{matplotlib}, \texttt{numpy} \citep{numpy}, \texttt{scipy} \citep{scipy}, \texttt{cmasher} \citep{cmasher} and \texttt{CoReCon} \citep{corecon}.

\section*{Data Availability}
All simulation data, including snapshots, group and subhalo catalogues, merger trees, and high time cadence Cartesian outputs are publicly available at \url{www.thesan-project.com} and thoroughly described in \citet{Thesan_data}, including the additional synthetic spectra developed for this work.

\section*{Author contributions}
We list here the authors contribution following the CRediT\footnote{\url{https://www.elsevier.com/researcher/author/policies-and-guidelines/credit-author-statement}} system. EG: conceptualization, methodology, software, formal analysis, validation, writing -- original draft, writing -- review and editing, visualization, supervision, project administration. VB: software, formal analysis, writing -- review and editing.

\bibliographystyle{mnras}
\bibliography{bibliography}

\appendix
\section{Public release of 300 additional sightlines and lightcone-like LOS}
\label{app:los_on_website}

\begin{table*}
    \caption{Properties added to and removed from the lightcone-like ray files released alongside with this paper, with respect to the ones available for the non-lightcone-like LOS files available at \url{www.thesan-project.com}. }
    \label{tab:los_lc_dsets}
    \centering
    \renewcommand{\arraystretch}{1.25}
    \begin{tabularx}{0.99\textwidth}{l|c|l}
         \textbf{Dataset} & \textbf{Units} & \textbf{Description} \\
         \hline
         Redshift & - & Redshift of each segment. Notice that this is the redshift the segment \textit{would have} if the LOS was extracted from a real lightcone (\ie without piecewise-constant approximation). To know the actual redshift, see the field `Snapshot'\\
         Snapshot & - & Snapshot from which each segment was extracted.\\
         RayDirections & - & Direction of each LOS. \\
         RayEndings & - & \textcolor{BrickRed}{removed} \\
    \end{tabularx}
    \renewcommand{\arraystretch}{0.8}
\end{table*}

Concurrently with this paper, we release the 300 additional sightlines produced at each snapshot between $5.5 \lesssim z \lesssim 7$ for \thesanone, as well as the lightcone-like LOS described in \ref{subsec:spectra}. All the data can be downloaded following the instructions at \url{www.thesan-project.com/data.html}. 

The 300 additional sightlines have the same data format as the original 150 LOS described in Section 3.10 of \citet{Thesan_data}. They are located at \texttt{Thesan/Thesan-1/postprocessing/los\_zdir}. In this directory, there is one file per snapshot containing all LOS. These files are named \texttt{rays\_zdir\_NNN.hdf5}, where \texttt{NNN} is the 0-left-padded snapshot number (starting from 54). 

The lightcone-like LOS are all contained in a single file located at \texttt{Thesan/Thesan-1/postprocessing/los\_lc/rays\_lc.hdf5}. They share the same format as the fixed-redshift LOS, but contain some additional datasets, reported in Table~\ref{tab:los_lc_dsets}.

\section{Impact of peculiar velocities}
\label{app:vpec}

\begin{figure}
    \includegraphics[width=\columnwidth]{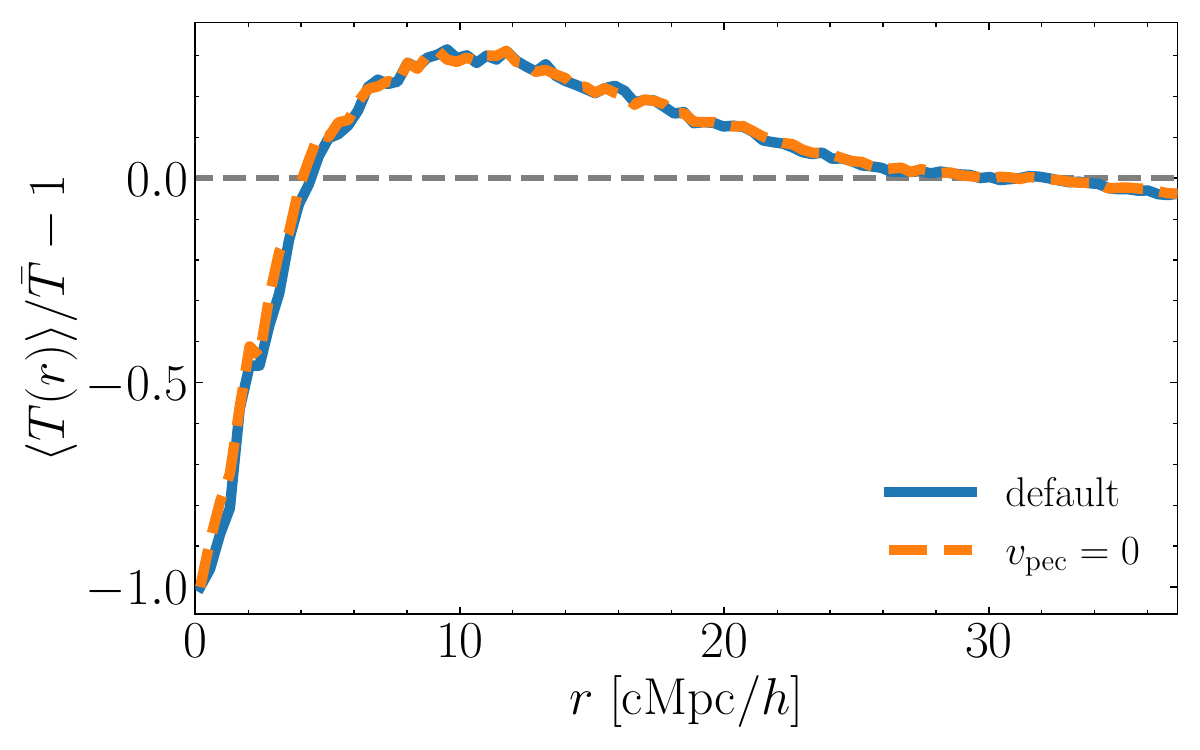}
    \caption{Impact of gas peculiar velocity on the \galacc. The solid line shows our fiducial spectra (\ie including peculiar velocities) while the dashed one is computed for spectra with artificially-static gas. }
    \label{fig:vpec}
\end{figure}

In Fig.~\ref{fig:vpec} we show the (vanishing) impact of including (fiducial, solid line) or ignoring (dashed line) peculiar velocities when computing the synthetic \Lya spectra. As it can be seen clearly, the impact is negligible.

\section{Impact of forcing the flux to be positive}
\label{app:noise_pos_flux}

\begin{figure}
    \includegraphics[width=\columnwidth]{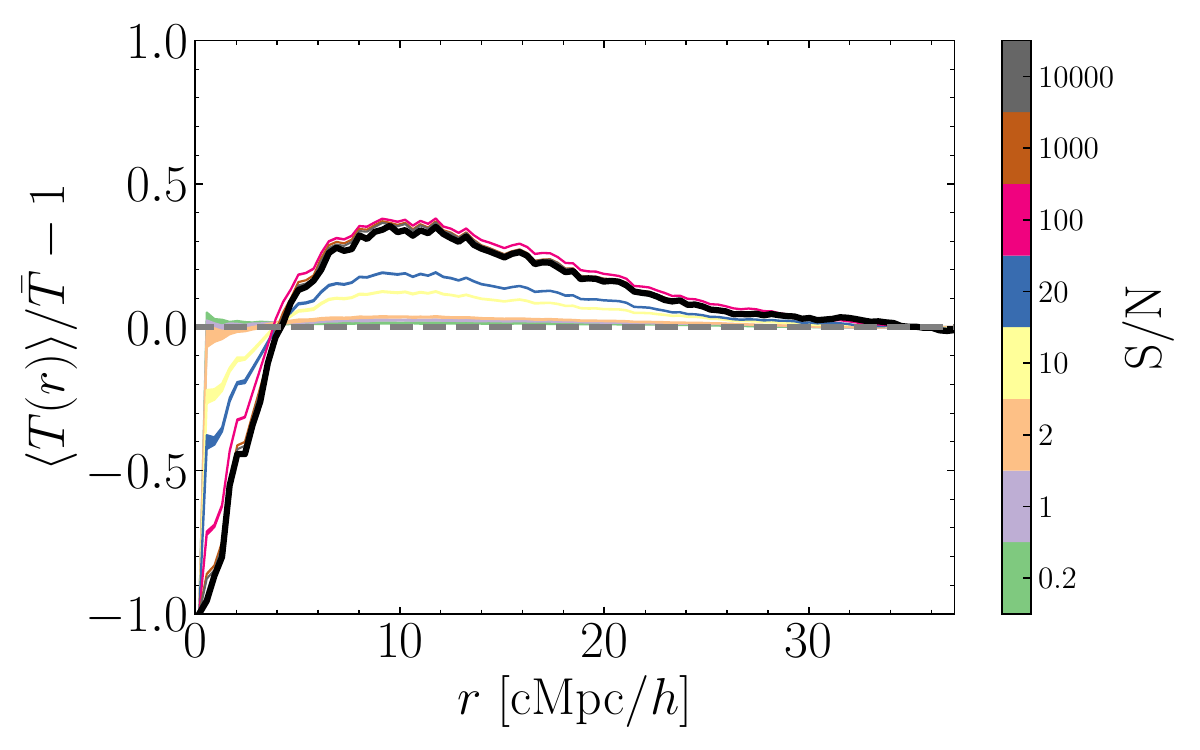}
    \caption{As Fig.~\ref{fig:corr_noise}, but now forcing the normalised flux to be positive (\ie zeroing pixels with negative flux). Unlike the previous case, very high S/N are required to recover the original signal. }
    \label{fig:corr_noise_positiveflux}
\end{figure}

In our investigation of the impact of noise on the \galacc in Sec.~\ref{subsec:noise_specres} we do not impose the flux in each pixel to be positive after the addition of noise. This choice stems from the consideration that errors in the quasar continuum estimation can lead to negative flux values. Here we investigate the impact of this choice. Interestingly, we find that forcing the flux to be positive makes the cross-correlation signal much less resilient to increasing noise in the spectrum, simultaneously changing the way it is affected by the noise. We show the impact of forcing the flux in each pixel to be non-negative (essentially by setting all negative flux values to zero) on the cross-correlation in Fig.~\ref{fig:corr_noise_positiveflux}. The difference with Fig.~\ref{fig:corr_noise} is striking. In this case, the noise does not simply broaden the range of recovered cross-correlation values around the intrinsic one, but instead suppresses the variations (both at small and intermediate scales) and eventually flattens the signal. Therefore, even in the case of moderate S/N (approximately 30\%), the recovered signal is far from the intrinsic one. Even more worryingly, the recovered signal appears similar to the intrinsic one at a later time (or, equivalently, larger ionized fraction), potentially preventing the ability to use this probe to pinpoint the completion of the reionization process. 
The reason of this behaviour change lies in the fact that imposing a positive flux introduces an asymmetry, since now noise can only increase the total flux, unlike in the previous case. Therefore, the noise does not average out when computing the cross-correlation. Instead, it effectively increases the flux regardless of the position of nearby galaxies, hence flattening the cross-correlation.

\section{Comparison of effective, median and middle redshift}
\label{app:zeff_zmed_zmid}

\begin{figure}
    \includegraphics[width=\columnwidth]{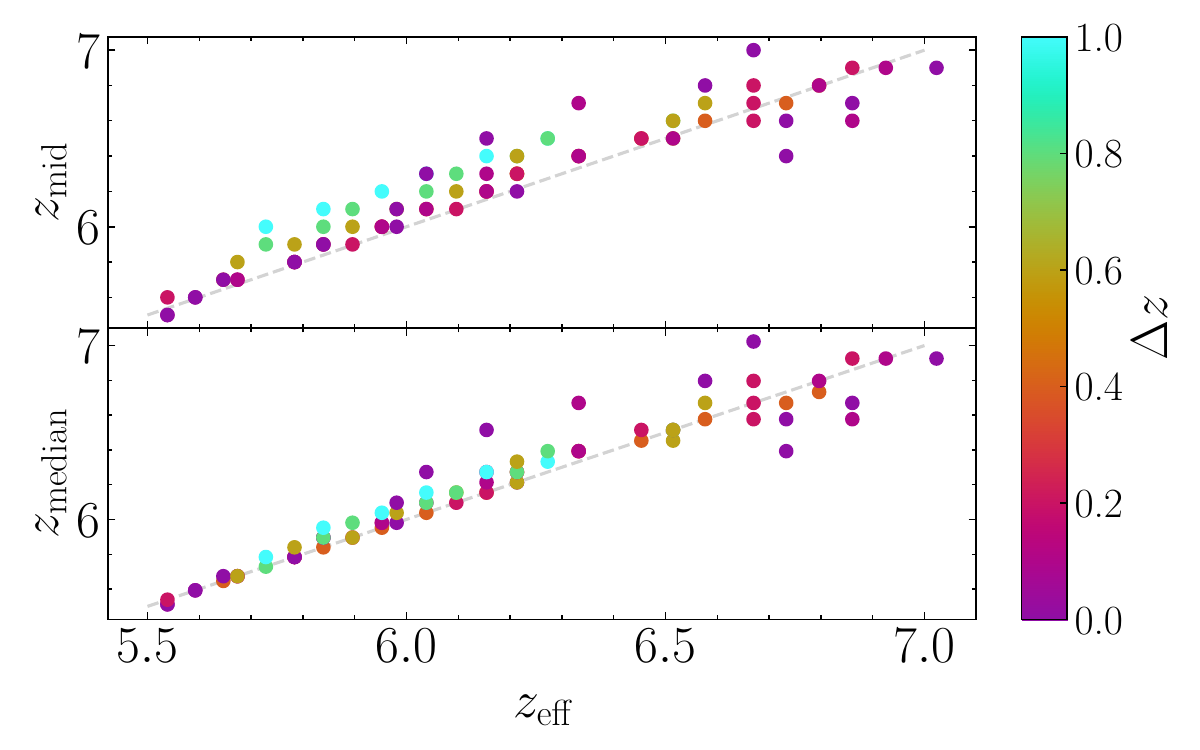}
    \caption{Comparison between the effective redshift ($z_\mathrm{eff}$, defined as the redshift of the fixed-redshift \galacc best-matching the lightcone-like one) and the central redshift of the window used ($z_\mathrm{mid}, $top panel) or the median redshift of the galaxies within the redshift window ($z_\mathrm{median}$, bottom panel). The colour of each point reflects the spectral window width $\Delta z$ employed to compute that specific lightcone-like cross correlation. We mark the one-to-one relation with a dashed grey line. $z_\mathrm{median}$ provides a better match to the actual $z_\mathrm{eff}$, but deviations are still found, especially at high redshift. Interestingly, $z_\mathrm{median}$ appears to fully capture the impact of large redshift windows. }
    \label{fig:corr_lightcone_effects_zeff_zmed_zcen}
\end{figure}

In Sec.~\ref{subsec:lightcone_effects} we have computed an effective redshift $z_\mathrm{eff}$ of the \galacc computed over a broad redshift window by identifying the closest \galacc computed at fixed redshift (or, equivalently, computed over a narrow redshift range). In Fig.~\ref{fig:corr_lightcone_effects_zeff_zmed_zcen} we show how such effective redshift tracks the central redshift of the window used ($z_\mathrm{mid}$, top panel) and the median redshift of all galaxies used ($z_\mathrm{median}$, bottom panel), for all combinations of $z_\mathrm{mid}$ and $\Delta z$ (the latter shown by the color of the points). In the bottom panel, points are much closer to the one-to-one line (dashed grey line), showing that $z_\mathrm{median}$ tracks much better the effective redshift of the cross-correlation computed from the lightcone-like LOS. 

Interestingly, $z_\mathrm{median}$ appears to almost-fully capture the impact of large redshift windows, as can be seen in the bottom panel of Fig.~\ref{fig:corr_lightcone_effects_zeff_zmed_zcen}, where all light blue/green points lie close to the one to one line (grey dashed line). This is not true for $z_\mathrm{mid}$. At $z_\mathrm{mid} \gtrsim 6.2$, the scatter around the $z_\mathrm{median} = z_\mathrm{eff}$ line increases, indicating that the effective redshift is a worse predictor of $z_\mathrm{eff}$ than at lower redshift. This is due to the increasing noise in the cross-correlation that makes the identification of an effective redshift more dependent on the noise (arising from the lower number of galaxies fulfilling our selection criteria (\ie $M_\mathrm{star} \geq 10^9 \, \Msun$) decreases at earlier and earlier times. We note that this would be true for any selection criteria based on galaxy mass or brightness.

\section{\galacc evolution fit}
\label{app:peak_evol_fit}

\begin{figure}
    \includegraphics[width=\columnwidth]{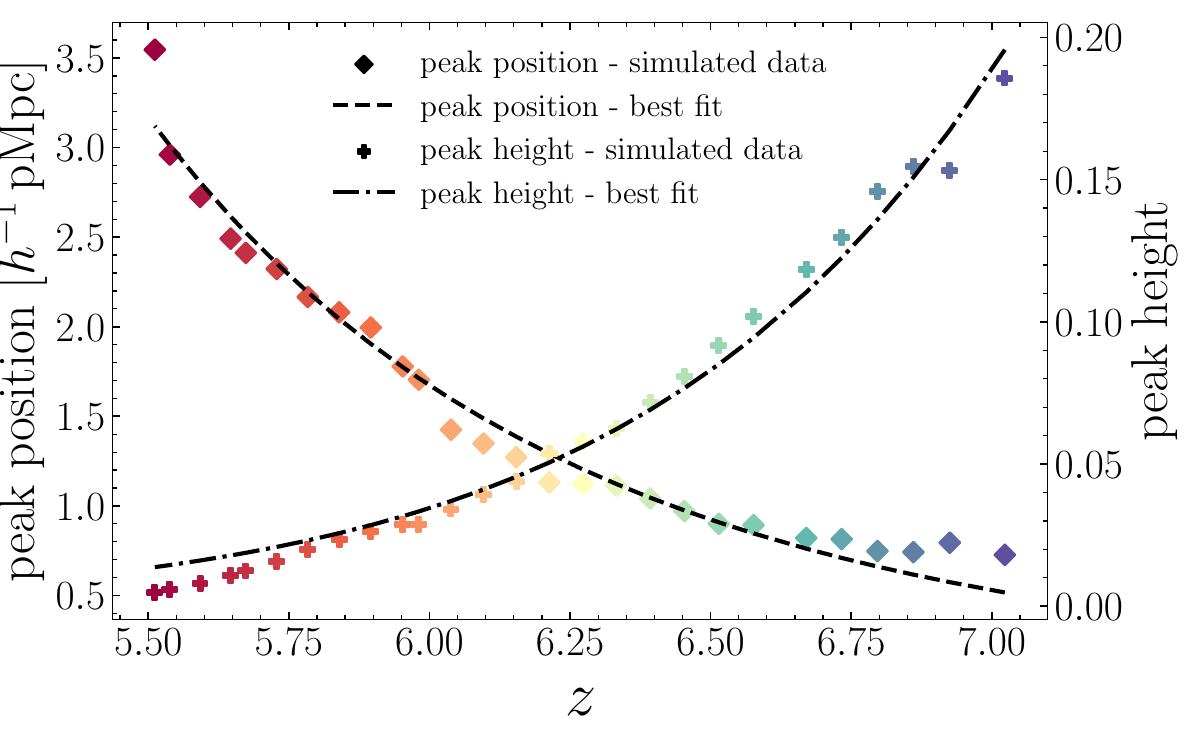}
    \caption{Evolution of the \galacc peak position (diamonds) and height (plus symbols) inferred from the \thesanone simulation. The best-fit evolution (following the functional form in Eq.~\ref{eq:fit_function}) is also shown using a dashed and dot-dashed line, respectively.}
    \label{fig:evol_fit}
\end{figure}

We report here the evolution of the amplitude and position of the peak in the \galacc computed from the \thesanone simulation. In order to identify them, we smooth the computed \galacc (using a running mean with window size 2 cMpc/$h$) to remove fluctuations and then identify the maximum of the curve as the peak of the cross-correlation. We have verified that the peak location and height are robust to the size of the window used in the smoothing procedure, and report their evolution in Fig.~\ref{fig:evol_fit} using diamond and plus symbols, respectively). We separately fit the peak position and height evolution with a function of the form \citep[in analogy to the mean free path evolution found by][]{Worseck+2014}:
\begin{equation}
\label{eq:fit_function}
    y = A \left( \frac{1+x}{6} \right)^\eta \, ,
\end{equation}
and report the best-fit evolution with a dashed and dot-dashed line in the Figure. Finally, we report the best-fitting parameter in Table~\ref{tab:best_fit_param}.

\begin{table}
    \caption{Best-fit parameters for the \galacc peak position and height evolution computed from the \thesanone simulation. The data are fit with the function in Eq.~\ref{eq:fit_function}.
    }
    \label{tab:best_fit_param}
    \centering
    \renewcommand{\arraystretch}{1.25}
    \begin{tabular}{l|c|c}
                        & \textbf{A} & $\mathbf{\eta}$ \\
         \hline
         peak position  & 6.32 & $-8.62$\\
         peak height    & 4.86 & 12.72\\
    \end{tabular}
    \renewcommand{\arraystretch}{0.8}
\end{table}

\label{lastpage}
\end{document}